\documentclass[a4paper,11pt]{article}
\usepackage{amsfonts,amsmath,amssymb,amscd}
\usepackage{appendix,marginnote,tikz,pgf,mathtools}
\usepackage{comment}
\usepackage{color}

\baselineskip=\normalbaselineskip

\def\ll{\left\langle}
\def\rr{\right\rangle}
\def\nn{\nonumber\\}

\def\i{\infty}
\def\k{\kappa}
\def\r{\lambda}
\def\s{\widetilde{t}_0}
\def\t{{t}_0}
\def\be{\begin{equation}}
\def\ee{\end{equation}}
\def\ben{\begin{equation*}}
\def\ena{\end{equation*}}
\def\bea#1\ena{\begin{align}#1\end{align}}
\def\bean#1\enan{\begin{align*}#1\end{align*}}

\def\p{\partial}

\begin{document}
\interfootnotelinepenalty=10000
\setcounter{footnote}{0}
\setcounter{tocdepth}{3}
\bigskip
\def\thefootnote{\arabic{footnote}}
\begin{titlepage}
\renewcommand{\thefootnote}{\fnsymbol{footnote}}
\begin{normalsize}
\begin{flushright}
\begin{tabular}{l}
APCTP Pre2020 - 007 \hspace{0.4cm}
\end{tabular}
\end{flushright}
\end{normalsize}

~~\\

\vspace*{0cm}
    \begin{Large}
       \begin{center}
         {Noncommutativity in two-matrix model extension of\\
          one-dimensional topological gravity}
       \end{center}
    \end{Large}
\vspace{0.7cm}

\begin{center}
Hisayoshi M\textsc{uraki}$^{1)}$\footnote[0]
            {
e-mail address : 
hisayoshi.muraki@apctp.org}

\vspace{0.7cm}

     $^{ 1)}$
     {\it Asia Pacific Center for Theoretical Physics, }\\
     {\it Pohang, 37673, Gyeongsangbuk-do, Korea}
\end{center}

\vspace{0.5cm}

\begin{abstract}
\noindent
One-dimensional topological gravity 
is defined as a Gaussian integral as its partition function.
The Gaussian integral supplies a toy model as a simpler version of one-matrix model that
is well known to provide a description of two-dimensional topological gravity.
The one-dimensional topological gravity inherits an integrable hierarchy structure
as with two-dimensional topological gravity, yet it is the Burgers hierarchy rather than 
the Korteweg--de Vries hierarchy.
Making use of this fact, an extension of the one-dimensional topological gravity
to an analogue of two-matrix model is investigated and the associated partition function
is shown to consist of a pair of partition functions of one-dimensional topological
gravity intertwined via the Moyal--Weyl product, which enables to provide
an explicit formula for its free energy. The extended system
shows a hierarchy structure interpreted as a noncommutative extension of the Burgers hierarchy.
The relation to noncommutative U(1) gauge theory is suggested.
\end{abstract}

\end{titlepage}

\tableofcontents

\section{Introduction}

One-dimensional topological gravity is formulated by one-dimensional Gaussian integral
as a toy model of one-matrix models of being well-known capable of describing
two-dimensional topological gravity on closed Riemann surfaces.
The two-dimensional quantum gravity, its matrix model descriptions
and their applications to statistical physics, string theory, enumerative geometry and so on
have been investigated in a variety of literature in both physics and mathematics, 
having originated from a vast amount of studies since the late 80's, for example,
\cite{David1988,Kostov1988,Distler1988,KPZ,Kazakov1989,Staudacher1989,Gross1990,BK1990,DS1990,GM1990,Ginsparg:1990}
and for more references, see \cite{Ginsparg1993,DiFrancesco1995}.

The partition function of the one-dimensional topological gravity shows
an integrable hierarchical structure called Burgers hierarchy and 
is subject to a set of Virasoro constraints.
This can be thought of as a miniature version of that 
the partition function of the two-dimensional topological gravity,
in other words, the partition function of one-matrix model, 
is subject to the KdV hierarchy \cite{D,Witten_91,Konst_92} as well as 
Virasoro constraints \cite{FKN_91,DVV_91}.
The hierarchy structure of one-matrix model acquires more richness
by extending to two-matrix models \cite{Gava1991,Daul1993,Staudacher1993,Eynard2003,Kazakov2003,Chekhov2006}.
This article investigates what kind of underlying structure appears
if the one-dimensional topological gravity is extended to its two-matrix model analogue.

The article is organized as follows. Section 2 starts with introducing
a Gaussian integral with a potential given by an infinite set of monomials
as a partition function of one-dimensional topological gravity.
It is reviewed that the partition function and the free energy
obey the Virasoro constraints, and a notion of genus expansion of free energy
makes sense by applying a saddle-point method to perform the integral.

Section 3 discusses identification of a recurrence relation for 
the free energy of one-dimensional topological gravity
with the Burgers hierarchy in some detail. The identification provides an implication of
the Cole--Hopf transform in the context of one-dimensional topological gravity. 
It is demonstrated how to find the free energy order by order
in the genus expansion making use of the hierarchy structure. 

Section 4 addresses the problem of an extension
of one-dimensional topological gravity to its two-matrix model analogue.
Based on the fact that the partition function of the one-dimensional topological gravity
obeys the Burgers hierarchy, the partition function of the extended model
is shown to be the Moyal--Weyl product of a pair of the
partition functions of one-dimensional topological gravity.
The noncommutative parameter turns out to be proportional to a coupling constant
describing interaction between two identical one-matrix models,
so that under the limit of decoupling the two systems the noncommutativity disappears
and the partition function is reduced to an ordinary product of each.
With the help of the Baker--Campbell--Hausdorff formula, which can be applied to
the Moyal--Weyl product, the free energy is shown to have genus expansions
and its explicit formula in terms of the free energy of one-dimensional gravity is obtained.
The associated hierarchy structure turns out to contain two sets of Burgers hierarchies.
Based on the formulation in terms of noncommutative Moya-Weyl product,
a noncommutative extension of the Burgers hierarchy rather naturally appears.

Section 5 clarifies the hierarchy structure arising
from the description based on the Moyal--Weyl product.
The partition function of the two-matrix model extension of one-dimensional
topological gravity is shown to be a tau-function of
the noncommutative Burgers hierarchy.
The relation to noncommutative U(1) gauge theory is discussed.
The final section summarizes the results and is mainly devoted to discussion, 
and appendices provide some supplemental information.

\section{One-dimensional topological gravity}

\subsection{Gaussian integral as partition function}

As reviewed briefly in this section,
the Gaussian integral shares many features of two-dimensional topological gravity, 
as well as matrix model. 
In this sense, it is considered as a toy model of two-dimensional gravity of 
matrix model description. 
Some recent developments on one-dimensional topological gravity
can also be found in \cite{Zhou1412,Zhang1904}, and 
supplemental analysis from the viewpoint of quasi-triviality 
of nonlinear partial differential equations has been made in \cite{DY1905}.

The partition function of the topological one-dimensional gravity is defined by
\be
\label{part.funct}
	Z^{1 D}(t)=\frac{1}{\sqrt{2 \pi} \lambda} \int^\infty_{-\infty} e^{\frac{1}{\lambda^{2}} S(x)} d x
\ee
with action
\be
	S(x)=-\frac{1}{2} x^{2}+\sum_{n=0}^{\infty} t_{n} \frac{x^{n+1}}{(n+1) !}
\ee
It appears as a Gaussian integral with an infinite set of monomial deformations as a potential.
Though the integral is not convergent in the presence of such a potential, 
it is regarded as a formal object and
its perturbative series in terms of the coupling constants is
to be understood as a formal expansion. 
The normalization factor of the partition function is understood by turning off all the couplings
\be
	Z^{1 D}(t=0)=\frac{1}{\sqrt{2 \pi} \lambda} \int_{-\infty}^\infty
	 e^{-\frac{x^2}{2\lambda^{2}}} d x=1
\ee

A formal expansion of the integrand in the coupling constants results in
\be
	Z^{1 D}(t)=\frac{1}{\sqrt{2 \pi} \lambda} 
	\int^\infty_{-\infty} e^{-\frac{x^2}{2\lambda^2}}
	\sum_{n_0,n_1,n_2,\cdots}
	\frac{1}{n_0!n_1!n_2!\cdots}
	\left(\frac{t_0}{\lambda^2}\frac{x^1}{1!}\right)^{n_0}
	\left(\frac{t_1}{\lambda^2}\frac{x^2}{2!}\right)^{n_1}
	\left(\frac{t_2}{\lambda^2}\frac{x^3}{3!}\right)^{n_2}
	\cdots d x
\ee
By virtue of a formula of the Gaussian integral
\be
	\frac{1}{\sqrt{2 \pi} \lambda} 
	\int^\infty_{-\infty} e^{-\frac{x^2}{2\lambda^2}}
	x^{n} dx
	=
	\begin{cases}
	0 &\quad (n={\rm odd})\\
	(n-1)!!\lambda^n &\quad (n={\rm even})
	\end{cases}
\ee
(where $(-1)!!\equiv1$) the partition function is evaluated as
\be
\label{part.func.eval}
	Z^{1 D}(t)= 
	\sum_{n_0,n_1,n_2,\cdots}
	\frac{\lambda^{n_0+2n_1+3n_2+\cdots \ }(-1+n_0+2n_1+3n_2+\cdots)!!}
	{n_0!n_1!n_2!\cdots \ 1!^{n_0}2!^{n_1}3!^{n_2}\cdots}
	\frac{t_0^{n_0}t_1^{n_1}t_2^{n_2}\cdots}{\lambda^{2n_0+2n_1+2n_2+\cdots}}
\ee
where the summation is taken over $n_0+2n_1+3n_2+\cdots$ being an even number.

The logarithm of the partition function defines the free energy
\be
	M^{1D}(t)=\log Z^{1 D}(t)
\ee
in other words, the partition function is given by the exponential of the free energy
\be
	Z^{1 D}(t)=e^{M^{1D}(t)}
\ee
The main focus of the paper is rather on $M^{1D}(t)$ than $Z^{1 D}(t)$.

\subsection{Virasoro constraints}

The partition function  $Z^{1 D}(t)$ is equipped with an infinity set of constraints
\be
\label{ViraConZ}
	L_m \, Z^{1 D}(t)=0
	\qquad
	(m\geq-1)
\ee
with
\be
\begin{aligned}
	L_{-1}&=\frac{t_0}{\lambda^2}+
	\sum_{n=1}^{\infty}\left(t_{n}-\delta_{n,1}\right)\frac{\p}{\p t_{n-1}}
	\\
	L_{0}&=1+\sum_{n=0}^{\infty}(n+1)\left(t_{n}-\delta_{n,1}\right)\frac{\p}{\p t_{n}}
	\\
	L_m&=\lambda^2 (m+1)!\frac{\p}{\p t_{m-1}}+
	\sum_{n=0}^{\infty}\frac{(m+n+1)!}{n !}\left(t_{n}-\delta_{n,1}\right)\frac{\p}{\p t_{m+n}}
	\qquad
	(m\geq1)
\end{aligned}
\ee
See appendix \ref{AppA} for its derivation. 
The operators form the following (part of the Virasoro) algebra
\be
	[L_m,L_n]=(m-n)L_{m+n}
	\qquad
	(m,n\geq-1)
\ee
The set of constraints can be phrased in terms of the free energy as follows.
\be
\begin{aligned}
\label{stringM}
	0
	&=\frac{t_0}{\lambda^2}+
	\sum_{n=1}^{\infty}\left(t_{n}-\delta_{n,1}\right)\frac{\p M^{1D}(t)}{\p t_{n-1}}
	\\
	0&=1+\sum_{n=0}^{\infty}(n+1)\left(t_{n}-\delta_{n,1}\right)\frac{\p M^{1D}(t)}{\p t_{n}}
	\\
	0&=\lambda^2 (m+1)!\frac{\p M^{1D}(t)}{\p t_{m-1}}+
	\sum_{n=0}^{\infty}\frac{(m+n+1)!}{n !}\left(t_{n}-\delta_{n,1}\right)\frac{\p M^{1D}(t)}{\p t_{m+n}}
	\qquad
	(m\geq1)
\end{aligned}
\ee

\subsection{Saddle point method and genus expansion}

Although the partition function of one-dimensional topological gravity
is simple enough to formally perform the integral as its 
result is already provided in \eqref{part.func.eval},
let us further explore the function from another viewpoint.
Eventually, there turns out to be the notion of ``genus expansion'' of the free energy.

The dominant contribution to the partition function \eqref{part.funct} comes from 
a saddle point defined by
\be
\label{saddlept}
	0=\frac{dS}{dx}(x)=-x+\sum_{n=0}^{\infty} t_{n} \frac{x^{n}}{n !}
\ee
Let $x_\infty$ denote a solution to the saddle point equation \eqref{saddlept}
\be
\label{def:xinf}
	x_\i=\sum_{n=0}^{\infty} t_{n} \frac{x_\i^{n}}{n !}
\ee

The Taylor expansion of the action $S(x)$ around the saddle point $x_\i$
gives an alternative expression of the partition function of form
\be
\label{part.funct.alt}
	Z^{1 D}(t)
	=\frac{e^{\frac{1}{\lambda^{2}} S(x_\i)}}{\sqrt{2 \pi} \lambda} \int 
	e^{
	-\frac{1-I_1}{2\lambda^2}y^2
	+\frac{1}{\lambda^{2}} \sum_{n=3}^\infty I_{n-1}\frac{y^n}{n!}} d y
\ee
where a change of variable $x\to y=x-x_\i$ is performed, and for $n\geq2$
\be
	\frac{d^n S}{dx^n}(x_\i)
	=-\delta_{n,2}+\sum_{m=0}^{\infty} t_{n-1+m} \frac{x^{m}_\i}{m!}
	=-\delta_{n,2}+I_{n-1}
\ee
with introducing a set of variables \cite{IZ,ACKM}
\be
	I_n=\sum_{m=0}^{\infty} t_{n+m} \frac{x^{m}_\i}{m !}
\ee
then in particular $I_0=x_\i$ and thus $I_n=\sum_{m=0}^{\infty} t_{n+m} \frac{I_0^{m}}{m !}$.
See appendix \ref{AppB} for some detailed relations between $\{t_k\}$ and $\{I_k\}$.
For later convenience it is worth mentioning that
\be
\label{dxdt0}
	\frac{\p x_\i}{\p t_0}=\frac{1}{1-I_1}
\ee
and thus
\be
\label{dIndt0inI}
	\frac{\p I_n}{\p t_0}=\frac{I_{n+1}}{1-I_1}
	\qquad (n\geq1)
\ee

Expanding the non-Gaussian part of \eqref{part.funct.alt} in power series in $I_k$,
performing the Gaussian integral yields
\be
\label{part.funct.alt.result}
	Z^{1 D}(t)	
	=
	\frac{e^{\frac{1}{\lambda^{2}} S(x_\i)}}{\sqrt{1-I_1}}
	\sum_{n_2,n_3,n_4,\cdots}
	\frac{I_{2}^{n_2}I_{3}^{n_3}I_{4}^{n_4}\cdots }
	{n_2!n_3!n_4!\cdots}\cdot
 	\frac{(1+3n_2+4n_3+5n_4+\cdots)!!}{3!^{n_2}4!^{n_3}5!^{n_4}\cdots}
	\left(\frac{\lambda}{\sqrt{1-I_1}}\right)^{n_2+2n_3+3n_4+\cdots}
\ee
with $3n_2+4n_3+5n_4+\cdots$ being an even number.
Note that the contribution from the summation 
to $\lambda$ expansion starts with something like $1+\mathcal{O}(\lambda^2)$.
Then, using the relation \eqref{dxdt0}, the free energy reads
\be
	M^{1D}(t)=\frac{S(x_\i)}{\lambda^{2}}
	+\frac12\log\left(\frac{\p x_\i}{\p t_0}\right)
	+(\text{correction of $\mathcal{O}(\lambda^2)$})
\ee

From the above mentioned observation, the free energy turns out to have
 an expansion according to the order of $\lambda$ in a form
\be
\label{FEGenExp}
	M^{1D}(t)=\sum_{g=0}^\infty \lambda^{2g-2}\,M^{1D}_{(g)}(t)
\ee
which can be thought of as ``genus'' expansion 
in the same manner as that of two-dimensional topological gravity.
Following the notation associated with the notion of genus expansion, $x_\i$ can be written as
\be
\label{rec:M0}
	\frac{x_\i^n}{n!}=\frac{\p M^{1D}_{(0)}(t)}{\p t_{n-1}}
	\qquad
	(n\geq1)
\ee
with 
\be
\label{freeene0}
	M^{1D}_{(0)}(t)=S(x_\i) 
\ee
and then the relation \eqref{rec:M0} can be fashionably phrased as a recurrence relation
\be
\label{recurre:M0}
	\frac{\p M^{1D}_{(0)}}{\p t_{n}}
	=\frac{1}{n+1}\frac{\p M^{1D}_{(0)}}{\p t_{0}}\frac{\p M^{1D}_{(0)}}{\p t_{n-1}}
\ee
The relation \eqref{dxdt0} is represented as
\be
\label{dxdt0inI}
	\frac{1}{1-I_1}=\frac{\p^2 M^{1D}_{(0)}}{\p {t_{0}}^2}
\ee
and then $\lambda^0$-part of the free energy is given by
\be
\label{freeene1}
	M^{1D}_{(1)}(t)=\frac12\log\left(\frac{\p^2 M^{1D}_{(0)}(t)}{\p {t_{0}}^2}\right)
\ee

The set of constraints \eqref{stringM} is as well decomposed according to powers in $\lambda$,
in particular the first relation of \eqref{stringM} is written as
\be
	\frac{\p M^{1D}_{(g)}(t)}{\p t_{0}}=t_0\,\delta_{g,0}+
	\sum_{n=0}^{\infty}t_{n+1}\frac{\p M^{1D}_{(g)}(t)}{\p t_{n}}
\ee
One can show that this is true for $g=0,1$
by direct evaluation making use of \eqref{def:xinf} and \eqref{rec:M0},
together with \eqref{freeene1}. 
This ensures that the free energy consistently has a $\lambda$-expansion of form
\eqref{FEGenExp}.

\section{Hierarchy structure}

\subsection{Burgers hierarchy and Cole--Hopf transformation}

The recurrence relation \eqref{recurre:M0} is found
for the lowest order in $\lambda$-expansion of the free energy $ M^{1D}$.
It would be natural to expect that there should be similar recurrence relations for
any order in $\lambda$-expansion. The answer takes the form
\be
\label{recurre:full}
		\frac{n+1}{\lambda^{2}} \frac{\partial M^{1D}}{\partial t_{n}}
	=\frac{\partial M^{1D}}{\partial t_{0}} \frac{\partial M^{1D}}{\partial t_{n-1}}
	+\frac{\partial^{2} M^{1D}}{\partial t_{0} \partial t_{n-1}}
\ee
which immediately follows from a relation
\be
\label{BZ}
	\frac{(n+1)!}{\lambda^{2n}}\frac{\p Z^{1D}}{\p t_{n}}
	=\frac{\p^{n+1}Z^{1D}}{\p t_0^{n+1}}
	=\frac{n!}{\lambda^{2n-2}}\frac{\p^2 Z^{1D}}{\p t_{0}\p t_{n-1}}
\ee
that is easily shown by the definition of the partition function \eqref{part.funct}.
For completeness, some consistency checks are provided in Appendix \ref{AppConsist}

The recurrence relation \eqref{recurre:full} has an underlying hierarchical structure,
which is transparent by rewriting it in a form
\be
	\frac{n+1}{\lambda^{2}} \frac{\partial M^{1D}}{\partial t_{n}}
	=\left(\frac{\p}{\p t_0}+\frac{\partial M^{1D}}{\partial t_{0}}\right)
	\frac{\p M^{1D}}{\p t_{n-1}}
\ee
and thus
\be
\label{recurre:full:BurHier}
	\frac{(n+1)!}{\lambda^{2n}}\, \frac{\partial M^{1D}}{\partial t_{n}}
	=\left(\frac{\p}{\p t_0}+\frac{\partial M^{1D}}{\partial t_{0}}\right)^n
	\frac{\p M^{1D}}{\p t_{0}}
\ee
This set of equations
coincides with the Burgers hierarchy formulated by a set of equations
\be
\label{BurgHirach}
 	\frac{\p u}{\p \tau_{n}}-\frac{\p}{\p x}\left(\frac{\p}{\p x}+u\right)^nu=0
	\qquad
	n=0,1,2,\dots
\ee
under identifications
\be
	x=t_0
	\qquad\quad
	\tau_n=\frac{\lambda^{2n}}{(n+1)!}t_n
	\qquad\quad
	u=\frac{\p M^{1D}}{\p t_{0}}
\ee
It is well-known that the Burgers hierarchy is linearized by the Cole--Hopf transformation
\be
\label{CHtransform}
	u=(\log\theta)_x
\ee
Noting that
\be
	\frac{\p u}{\p \tau_{n}}
	=\frac{\p}{\p x}\left(\frac{1}{\theta}\frac{\p\theta}{\p\tau_{n}}\right)
	\qquad\qquad
	\left(\frac{\p}{\p x}+u\right)^nu
	=\frac{1}{\theta}\frac{\p^{n+1}\theta}{\p x^{n+1}}
\ee
the Burgers hierarchy is equivalent to a set of linear differential equations
\be
\label{BTheta}
	\frac{\p\theta}{\p\tau_{n}}=\frac{\p^{n+1}\theta}{\p x^{n+1}}
\ee
which is nothing but the relation \eqref{BZ} under identification $\theta=Z^{1D}$.
Some exact solutions to this linear differential equation
for specific situations are examined in the next subsection.

Using the property \eqref{BZ}, 
the set of Virasoro constraints \eqref{ViraConZ} can be written as
\be
	L_m\, Z^{1 D}(t)
	=\lambda^{2(m+1)}\,\frac{\p^{m+1}}{\p t_0^{m+1}}\cdot L_{-1}\, Z^{1 D}(t)
	\qquad
	(m\geq0)
\ee
so that it can reduce to a single condition $L_{-1}\, Z^{1 D}(t)=0$ (string equation).

\subsection{Exact solutions under special conditions \label{AppExact}}

In this subsection, for completeness, let it provide
some checks if the partition function
is actually a solution to the partial differential equation
under some specifically simpler setting.

\subsubsection{Burgers equation: Heat-kernel method}

Let all the couplings turned off except $t_0$ and $t_1$:
The partition function reduces to the simplest Gaussian integral resulting in
\be
\label{PFBurgers}
	Z^{1 D}(t_0,t_1)
	=\frac{1}{\sqrt{2 \pi} \lambda} \int^\infty_{-\infty} 
	e^{\frac{1}{\lambda^{2}} \left(-\frac{1-t_1}{2} x^{2}+t_0x\right)} d x
	=\frac{e^{\frac{t_0^2}{2\lambda^{2}(1-t_1)}}}{\sqrt{1-t_1}}
\ee
and then the free energy is exactly calculated as
\be
	M^{1 D}(t_0,t_1)=\frac{1}{\lambda^{2}}\,\frac{t_0^2}{2(1-t_1)}
	+\frac12\log\left(\frac{1}{1-t_1}\right)
\ee
in particular $M^{1 D}(t_0,t_1=0)=\frac{t_0^2}{2\lambda^{2}}$
so that $\frac{\p M^{1 D}}{\p t_0}(t_0,t_1=0)=\frac{t_0}{\lambda^{2}}$.

The Burgers equation \cite{Burgers}
\be
 	\frac{\p u}{\p \tau}-\frac{\p}{\p x}\left(\frac{\p}{\p x}+u\right)u=0
\ee
shortly
\be
\label{BurgersEq}
	u_{\tau}=u_{xx}+2uu_x
\ee
is equivalent to the one-dimensional heat equation
\be
	\frac{\p\theta}{\p\tau}=\frac{\p^{2}\theta}{\p x^{2}}
\ee
via the Cole--Hopf transformation \cite{Cole,Hopf}: $u=(\log\theta)_x$.
The heat equation is solved by the heat-kernel method putting
\be
	\theta(x,\tau)=\int dx'\, G(x-x',\tau)\theta(x',0)
\ee
with the heat-kernel and the initial condition being
\be
	G(x-x',\tau)=\frac{1}{\sqrt{4\pi\tau}}e^{-\frac{(x-x')^2}{4\tau}}
	\qquad\quad
	\theta(x',0)=e^{\int^{x'} dx''u(x'',0)}
\ee
respectively, where a direct evaluation shows
\be
	\left(\frac{\p}{\p \tau}-\frac{\p^2}{\p x^2}\right)G(x-x',\tau)=0
\ee
Performing the integration under initial condition $u(x,0)=\frac{x}{\lambda^2}$ yields
\be
	\theta(x,\tau)
	=\int dx'\, \frac{1}{\sqrt{4\pi\tau}}
	e^{-\frac{(x-x')^2}{4\tau}+\frac{x'^2}{2\lambda^2}}
	=\sqrt{\frac{\lambda^2}{\lambda^2-2\tau}}
	e^{-\frac{x^2}{4\tau}\left(1-\frac{\lambda^2}{\lambda^2-2\tau}\right)}
\ee
Substituting $\tau\to \frac{\lambda^2}{2}t_1$ and $x\to t_0$
reproduces \eqref{PFBurgers}.

\subsubsection{Burgers equation: ($G'/G$)-expansion method}

A technique for generating a part of exact solutions describing propagating waves
governed by nonlinear partial differential equations is 
the method called ($G'/G$)-expansion \cite{WLZ}.
Although a solution considered in this article will turn out
to be out of the range covered by the method, 
it may be worth explaining what is its obstruction.
A brief sketch of the method is given in appendix \ref{AppC}.

The method puts ansatz that 
the solution shall be the form of a (Laurent) polynomial of ($G'/G$):
\be
	u(x,\tau)=u(\xi)=\sum_{n=N_{\rm min}}^{N_{\rm max}} a_n\left(\frac{G'}{G}\right)^n
	\qquad\quad
	G=G(\xi)
\ee
with
\be
	\xi=x-c\tau
	\qquad\quad
	G'=\frac{d G}{d \xi}
\ee
assuming $G$ should solve the following ordinary differential equation
\be
	G''+rG'+sG=0
\ee
Under the ansatz, the Burgers equation $u_{\tau}=u_{xx}+(u^2)_x$ 
reduces to an ordinary differential equation
\be
\label{BurgEq:Ord}
	-cu'=u''+(u^2)'
\ee
which can be integrated as
\be
\label{BurgEq:Int}
	C=cu+u'+u^2
\ee
where $C$ is the integration constant.
Requiring the highest order in $(G'/G)$ shall be cancelled between $u'$ and $u^2$,
implies $N_{\rm max}+1=2N_{\rm max}$ and thus $N_{\rm max}=1$.
Applying a similar argument on $N_{\rm min}$ to the lowest order in $(G'/G)$ 
is implying $N_{\rm min}=0$:
\be
	u(x,\tau)=u(\xi)= a_1\left(\frac{G'}{G}\right)+a_0
\ee
Plugging this ansatz into \eqref{BurgEq:Int} (equivalently \eqref{BurgEq:Ord})
results in a set of simultaneous equations for the parameters
\be
	a_1=1
	\qquad\quad
	c=s+r-2a_0
	\qquad\quad
	C=(s+r-a_0)a_0
\ee
and one should chose a specific value of parameters consistently with initial conditions.

Coming back to the interest of the current article,
imposing the initial condition $u(x,0)=\frac{x}{\lambda^2}$ implies
\be
	\left.\frac{G'}{G}\right|_{\tau=0}=\frac{x}{\lambda^2}-a_0
\ee
so that
\be
	\frac{G'}{G}=\frac{\xi}{\lambda^2}-a_0
\ee
and therefore
\be
	G''+a_0G'-\frac{1}{\lambda^2}G
	=-\frac{\xi}{\lambda^2}G'
\ee
which is out of ansatz. This observation shows that the solution to Burgers equation
with the initial condition $u(x,0)=\frac{x}{\lambda^2}$
would not be written in the form of a closed form as a propagating wave
assumed by the ansatz.

\subsubsection{Sharma--Tasso--Olver equation}

The case where all the couplings except $t_0$ and $t_2$ are turned off
is associated with the Sharma--Tasso--Olver equation \cite{ST,Olver}:
\be
 	\frac{\p u}{\p \tau}-\frac{\p}{\p x}\left(\frac{\p}{\p x}+u\right)^2u=0
\ee
shortly
\be
\label{STOeq}
 	u_{\tau}=
	u_{xxx}+3(uu_x)_{x}+3u^2u_x
\ee
Correspondingly the partition function \eqref{part.func.eval} reduces to
\be
	Z^{1 D}(t_0,t_2)
	= \sum_{n_0,n_2}\frac{1}{3!^{n_2}}
	\frac{(-1+n_0+3n_2)!!}{n_0!\,n_2!}
	\frac{t_0^{n_0}t_2^{n_2}}{\lambda^{n_0-n_2}}
	=\sum_{m,n}\frac{(-1+m+3n)!!}{m!\,n!}\frac{x^{m}\tau^n}{\lambda^{m+3n}}
	=:\theta
\ee
with
\be
	x=t_0
	\qquad\quad
	\tau=\lambda^{4}\frac{t_2}{3!}
	\qquad\quad
	u=\frac{\p M^{1D}}{\p t_{0}}
\ee
The function $\theta$ satisfies
\be
	\frac{\p \theta}{\p \tau}
	=\sum_{m,n}\frac{(2+m+3n)!!}{m!\,n!}\frac{x^{m}\tau^{n}}{\lambda^{m+3n+3}}
	=\frac{\p^3 \theta}{\p x^3}
\ee
which is equivalent to the Sharma--Tasso--Olver equation via $u=(\log \theta)_x$.

\subsection{Higher order in genus expansion}

The first two lowest orders in $\lambda$ expansion of free energy $M^{1D}$ are 
somewhat trivially given by \eqref{freeene0} and \eqref{freeene1}
because of \eqref{part.funct.alt.result}.
Non-trivial information about the free energy can be extracted 
from the hierarchy structure \eqref{recurre:full} by looking into the next order:
\be
\label{recurre:order2}
	(n+1) \frac{\partial M^{1D}_{(2)}}{\partial t_{n}}
	=\frac{\partial M^{1D}_{(0)}}{\partial t_{0}} \frac{\partial M^{1D}_{(2)}}{\partial t_{n-1}}
	+\frac{\partial M^{1D}_{(2)}}{\partial t_{0}} \frac{\partial M^{1D}_{(0)}}{\partial t_{n-1}}
	+\frac{\partial M^{1D}_{(1)}}{\partial t_{0}} \frac{\partial M^{1D}_{(1)}}{\partial t_{n-1}}
	+\frac{\partial^{2} M^{1D}_{(1)}}{\partial t_{0} \partial t_{n-1}}
\ee
from which the $\lambda^2$-part of the free energy can be specified as
\be
\label{freeene2}
	M^{1D}_{(2)}
	=\left(\frac{\p^2 M^{1D}_{(0)}}{\p {t_{0}}^2}\right)^{-1}
	\left(\frac{\text{1}}{4}\frac{\p^2 M^{1D}_{(1)}}{\p {t_{0}^2}}
	-\frac{\text{1}}{6}\left(\frac{\p M^{1D}_{(1)}}{\p t_{0}}\right)^2\right)
\ee
A straightforward evaluation shows that $M^{1D}_{(2)}$ given by \eqref{freeene2} satisfies 
the relation \eqref{recurre:order2}, using the recurrence relations for $M^{1D}_{(0)}$ and $M^{1D}_{(1)}$.
This result reproduces the expression given in \cite{Zhang1904}:
\be
	M^{1D}_{(2)}
	=\frac{5}{24} \frac{I_{2}^{2}}{\left(1-I_{1}\right)^{3}}+\frac{1}{8} \frac{I_{3}}{\left(1-I_{1}\right)^{2}}
\ee
by noting \eqref{dIndt0inI} and \eqref{dxdt0inI}, together with \eqref{freeene1}.

Higher order in $\lambda$-expansion will involve more terms, but
still it is manageable to proceed to finding the higher order part of the free energy.
Let us demonstrate it here taking $M^{1D}_{(3)}$ and $M^{1D}_{(4)}$ as example.
From the above mentioned observation and dimensional analysis, one may put ansatz
\be
\begin{aligned}
\label{freeene3}
	M^{1D}_{(3)}
	&=\left(\frac{\p^2 M^{1D}_{(0)}}{\p {t_{0}}^2}\right)^{-1}
	\left\{a_1\frac{\p^2 M^{1D}_{(2)}}{\p {t_{0}^2}}
	+a_2\frac{\p M^{1D}_{(1)}}{\p t_{0}}\frac{\p M^{1D}_{(2)}}{\p t_{0}}\right\}
	\\&\quad
	+\left(\frac{\p^2 M^{1D}_{(0)}}{\p {t_{0}}^2}\right)^{-2}
	\left\{b_1\left(\frac{\p^2 M^{1D}_{(1)}}{\p t_{0}^2}\right)^2
	+b_2\frac{\p^2 M^{1D}_{(1)}}{\p t_{0}^2}\left(\frac{\p M^{1D}_{(1)}}{\p t_{0}}\right)^2
	+b_3\left(\frac{\p M^{1D}_{(1)}}{\p t_{0}}\right)^4\right\}
\end{aligned}
\ee
The coefficients are determined as
\be
	a_1=\frac{1}{6}
	\qquad
	a_2=-\frac{4}{9}
	\qquad
	b_1=-\frac{1}{9}
	\qquad
	b_2=\frac{5}{27}
	\qquad
	b_3=-\frac{2}{27}
\ee
by requiring the recurrence relation \eqref{recurre:full} at 
the corresponding order in $\lambda$:
\be
	(n+1) \frac{\partial M_{(3)}^{1D}}{\partial t_{n}}
	=\frac{\partial M_{(0)}^{1D}}{\partial t_{0}} \frac{\partial M_{(3)}^{1D}}{\partial t_{n-1}}
	+\frac{\partial M_{(3)}^{1D}}{\partial t_{0}} \frac{\partial M_{(0)}^{1D}}{\partial t_{n-1}}
	+\frac{\partial^{2} M_{(2)}^{1D}}{\partial t_{0} \partial t_{n-1}}
	+\frac{\partial M_{(1)}^{1D}}{\partial t_{0}} \frac{\partial M_{(2)}^{1D}}{\partial t_{n-1}}
	+\frac{\partial M_{(2)}^{1D}}{\partial t_{0}} \frac{\partial M_{(1)}^{1D}}{\partial t_{n-1}}
\ee
The next order appears more cumbersome, but one may still be able to put ansatz
and find a solution (to reduce the notational complexity, the prime stands for 
the derivative with respect to $t_0$)
\be
\begin{aligned}
\label{freeene4}
	M^{1D}_{(4)}
	&=\left({M''_{(0)}}\right)^{-1}
	\left\{\frac{1}{8}M''_{(3)}-\frac{1}{2}M'_{(1)}M'_{(3)}
	-\frac{2}{5}\left(M'_{(2)}\right)^2\right\}
	\\&
	+\left({M''_{(0)}}\right)^{-2}
	\left\{
	-\frac{5}{24}M''_{(1)}M''_{(2)}
	+\frac{7}{36}\left(M'_{(1)}\right)^2M''_{(2)}
	-\frac{82}{135}\left(M'_{(1)}\right)^3M'_{(2)}
	+\frac{31}{45}M'_{(1)}M''_{(1)}M'_{(2)}\right\}
	\\&
	+\left(M''_{(0)}\right)^{-3}
	\left\{
	\frac{7}{72}\left(M''_{(1)}\right)^3
	-\frac{13}{45}\left(M''_{(1)}\right)^2
		\left(M'_{(1)}\right)^2
	+\frac{217}{810}M''_{(1)}
		\left(M'_{(1)}\right)^4
	-\frac{32}{405}\left(M'_{(1)}\right)^6\right\}
\end{aligned}
\ee
which satisfies
\be
	(n+1) \frac{\partial M_{(4)}^{1D}}{\partial t_{n}}
	=\frac{\partial M_{(0)}^{1D}}{\partial t_{0}} \frac{\partial M_{(4)}^{1D}}{\partial t_{n-1}}
	+\frac{\partial M_{(4)}^{1D}}{\partial t_{0}} \frac{\partial M_{(0)}^{1D}}{\partial t_{n-1}}
	+\frac{\partial^{2} M_{(3)}^{1D}}{\partial t_{0} \partial t_{n-1}}
	+\frac{\partial M_{(1)}^{1D}}{\partial t_{0}} \frac{\partial M_{(3)}^{1D}}{\partial t_{n-1}}
	+\frac{\partial M_{(2)}^{1D}}{\partial t_{0}} \frac{\partial M_{(2)}^{1D}}{\partial t_{n-1}}
	+\frac{\partial M_{(3)}^{1D}}{\partial t_{0}} \frac{\partial M_{(1)}^{1D}}{\partial t_{n-1}}
\ee
This demonstrates how to put ansatz and how to find out the free energies order by order,
and how useful to formulate the system in terms of the
recurrence relation together with $\lambda$-expansion.
Another systematic approach for finding out $M_{(n)}^{1D}$
using a set of variable $\{I_n\}$ is found in \cite{Zhang1904}.

\section{Two-matrix model extension}

\subsection{Partition function and Moyal--Weyl product}

Let it move on to considering an extension of one-dimensional topological gravity.
An extension to two-matrix model analogue is one of straightforward possibilities.
A partition function for the analogue of two-matrix model can be defined by
\be
\label{TwommPF}
	{Z}^{\rm TM}(t,\widetilde{t},\k)
	=\frac{1}{{2 \pi}\lambda^2} 
	\int^\infty_{-\infty} \int^\infty_{-\infty} 
	e^{\frac{1}{\lambda^{2}} S(x)+\frac{1}{\lambda^{2}} \widetilde{S}(y)
	+\k\frac{xy}{\lambda^2}} dxdy
\ee
with
\be
	S(x)=-\frac{1}{2} x^{2}+\sum_{n=0}^{\infty} t_{n} \frac{x^{n+1}}{(n+1) !}
	\qquad\quad
	\widetilde{S}(y)
	=-\frac{1}{2} y^{2}+\sum_{n=0}^{\infty} \widetilde{t}_{n} \frac{y^{n+1}}{(n+1) !}
\ee
The interaction term between a pair of one-matrix model
under consideration in this and the subsequent subsections
is of type $xy$, whose matrix counterpart is considered as the first
nontrivial interacting two-matrix models, for example \cite{Eynard2003,Kazakov2003}.
From this view point, the partition function \eqref{TwommPF} shall be
regarded as a simpler toy model of interacting two-matrix model.

The normalization factor of the partition function is understood 
when all the couplings are turned off:
\bea
	{Z}^{\rm TM}(t=0,\widetilde{t}=0,\k)
	&=\frac{1}{{2 \pi}\lambda^2} \int^\infty_{-\infty} \int^\infty_{-\infty} 
	e^{-\frac{x^2}{2\lambda^{2}} }
	e^{-\frac{y^2}{2\lambda^{2}} }\sum_{n=0}^\infty \frac1{n!}
	\left(\k\frac{xy}{\lambda^2}\right)^n dxdy
	\nn
	&=
	\sum_{n=0}^\infty \frac1{n!}
	\left(\frac{\k}{\lambda^2}\right)^n
	\left(\frac{1}{\sqrt{2 \pi} \lambda} \int^\infty_{-\infty} 
	e^{-\frac{x^2}{2\lambda^{2}} }x^ndx\right)
	\left(\frac{1}{\sqrt{2 \pi} \lambda} \int^\infty_{-\infty} 
	e^{-\frac{y^2}{2\lambda^{2}} }y^ndy\right)
	\nn
	&=\left(
	1+\sum_{n=1}^\infty \frac{(2n-1)!!}{(2n)!!}\,\k^{2n}\right)=\frac{1}{\sqrt{1-\k^2}}
\ena
where the factor involving a summation coincides with the Taylor series of
$1/\sqrt{1-\k^2}$ around $\k=0$.
When one of the sets of couplings, say $\widetilde{t}_n$, is turned off,
the partition function is reduced to the one for one-dimensional topological gravity
\be
	{Z}^{\rm TM}(t,\widetilde{t}=0,\k)
	=Z^{1 D}\left(t_0,t_1+\k^2,t_{i\geq2}\right)
\ee

Retaining all the coupling constants,
the integral can be written in a form of a product
defined on a noncommutative plane $(\t,\s)$ owing to the fact that
the one-dimensional topological gravity is subject to the Burgers hierarchy:
\bea
	{Z}^{\rm TM}(t,\widetilde{t},\k)
	&=\sum_{n=0}^\infty \frac1{n!}
	\left(\frac{\k}{\lambda^2}\right)^n
	\left(\frac{1}{\sqrt{2 \pi} \lambda} \int^\infty_{-\infty} 
	e^{\frac{1}{\lambda^2}S(x)}x^ndx\right)
	\left(\frac{1}{\sqrt{2 \pi} \lambda} \int^\infty_{-\infty} 
	e^{\frac{1}{\lambda^{2}}\widetilde{S}(y)}y^ndy\right)
	\\
	&=\sum_{n=0}^\infty \frac1{n!}
	\left(\frac{\k}{\lambda^2}\right)^n
	\left(n!\lambda^2\frac{\p}{\p t_{n-1}}Z^{1 D}(t)\right)
	\left(n!\lambda^2\frac{\p}{\p \widetilde{t}_{n-1}}Z^{1 D}(\widetilde{t})\right)
	\\
	&=\sum_{n=0}^\infty \frac1{n!}
	\left(\k\lambda^2\right)^n
	\left(\frac{\p^{n}}{\p t_0^{n}}Z^{1 D}(t)\right)
	\left(\frac{\p^{n}}{\p \widetilde{t}_0^{n}}Z^{1 D}(\widetilde{t})\right)
	\\
	&=Z^{1 D}(t)\ast Z^{1 D}(\widetilde{t})
\ena
where $\ast$ stands for the noncommutative product known as 
the Moyal--Weyl product defined by
\be
\label{MWprod}
	f(t,\widetilde{t})\ast g(t,\widetilde{t})
	=f(t,\widetilde{t})\,e^{\frac{i\theta}{2}\left(
	\overset{\leftarrow}{\frac{\p}{\p t_0}}
	\overset{\rightarrow}{\frac{\p}{\p \widetilde{t}_0}}
	-\overset{\leftarrow}{\frac{\p}{\p \widetilde{t}_0}}
	\overset{\rightarrow}{\frac{\p}{\p t_0}}\right)}
	g(t,\widetilde{t})
\ee
with the noncommutative parameter\footnote{
In most cases, the noncommutative parameter $\theta$ is 
to be real $(\overline{\theta}=\theta)$,
to ensure the convention on complex conjugation is to be
$\overline{f\ast g}=\overline{g}\ast \overline{f}$. 
For example, the relation $[x,y]_\ast=i\theta$ makes sense
under this convention:
$
\overline{[x,y]_\ast}=\overline{x\ast y}-\overline{y \ast x}
=y \ast x-x \ast y=-[x,y]_\ast=-i\theta=\overline{i\theta}
$.
Whereas, in the case under consideration with $\theta$ 
being pure imaginary $(\overline{\theta}=-\theta)$,
the convention on complex conjugation is to be understood as
$\overline{f\ast g}=\overline{f}\ast \overline{g}$.
Under this convention, for example, the relation $[x,y]_\ast=i\theta$ makes sense, i.e.
$
\overline{[x,y]_\ast}=\overline{x\ast y}-\overline{y \ast x}
=x \ast y-y \ast x=[x,y]_\ast=i\theta=\overline{i\theta}
$.
}
\be
	\theta=-2i\k\lambda^2
\ee
Note that the Moyal--Weyl product is associative
\be
	\left(f(t,\widetilde{t})\ast g(t,\widetilde{t})\right)\ast h(t,\widetilde{t})
	=f(t,\widetilde{t})\ast \left(g(t,\widetilde{t})\ast h(t,\widetilde{t})\right)
\ee
but noncommutative, in general,
\be
	f(t,\widetilde{t})\ast g(t,\widetilde{t})
	\neq g(t,\widetilde{t})\ast f(t,\widetilde{t})
\ee
e.g. $f(t,\widetilde{t})=\t,\ g(t,\widetilde{t})=\s$
\be
	\t \ast \s =\t\s +\frac{i\theta}{2}
\ee
while
\be
	\s \ast \t =\t\s -\frac{i\theta}{2}\neq \t \ast \s
\ee
so that the $\theta$ accounts for the noncommutativity between $\t$ and $\s$
in the following sense
\be
\label{NCrelation}
	\left[\t,\s\right]_\ast=i\theta
\ee
where the commutator is defined by
\be
	\left[f,g\right]_\ast=f\ast g-g\ast f
\ee
As $\k\to0$ the noncommutativity disappears,
that accounts for that the system is reduced to a pair of copies of
one-dimensional topological gravity under the limit.

Let $M^{\rm TM}_\ast(t,\widetilde{t},\k)$ denote
 the ``free energy'' of two-matrix analogue
\be
\label{WeylExp}
	{Z}^{\rm TM}(t,\widetilde{t},\k)
	={\rm Exp}_\ast\left(M^{\rm TM}_\ast(t,\widetilde{t},\k)\right)
\ee
Here ${\rm Exp}_\ast$ is a map,
on any smooth function $X=X(t,\widetilde{t})$, whose action is given by
\be
	{\rm Exp}_\ast\left(X\right)
	=1+X+\frac{1}{2!}X\ast X+\frac{1}{3!}X\ast X\ast X+\dots
	=\sum_{n=0}^\infty\frac{1}{n!}X^{\ast (n)}
\ee
with $X^{\ast(n)}$ being recursively defined by
\be
	X^{\ast(n)}:=X\ast X^{\ast (n-1)}
\ee
which makes sense as the $\ast$-product is associative.
Note that the deviation from the ordinary exponential  purely arises from
the noncommutative parameter
\be
	{\rm Exp}_\ast\left(X\right)-{\rm exp}\left({X}\right)
	=\mathcal{O}(\theta)
\ee
It is convenient to introduce the (formal) inverse function ${\rm Log}_\ast$ by
\be
	X={\rm Log}_\ast\left({\rm Exp}_\ast\left(X\right)\right)
\ee

Then, following the exponential/logarithmic function relevant to the
Moyal--Weyl product, and the Baker--Campbell--Hausdolff formula
(for simplicity $M:=M^{1D}(t)$ and $\widetilde{M}:=M^{1D}(\widetilde{t})$),
the ``free energy'' $M^{\rm TM}_\ast(t,\widetilde{t},\k)$ can be explicitly given 
in terms of $M$ and $\widetilde{M}$:
\be
\label{BCHFreeEnerg}
\begin{aligned}
	M^{\rm TM}_\ast(t,\widetilde{t},\k)
	&=M+\widetilde{M}
	+\frac12\left[M,\widetilde{M}\right]_\ast\\
	&\qquad
	+\frac{1}{12}\left(
	\left[M,\left[M,\widetilde{M}\right]_\ast\right]_\ast
	+\left[\widetilde{M},\left[\widetilde{M},M\right]_\ast\right]_\ast
	\right)\\
	&\qquad
	-\frac{1}{24}
	\left[\widetilde{M},\left[M,\left[M,\widetilde{M}\right]_\ast\right]_\ast\right]_\ast
	\\
	&\qquad-\frac{1}{720}\left(
		\left[\widetilde{M},\left[\widetilde{M},\left[\widetilde{M},\left[\widetilde{M},M\right]_\ast\right]_\ast\right]_\ast\right]_\ast
		+\left[M,\left[M,\left[M,\left[M,\widetilde{M}\right]_\ast\right]_\ast\right]_\ast\right]_\ast\right)\\
	&\qquad+\frac{1}{360}\left(
		\left[M,\left[\widetilde{M},\left[\widetilde{M},\left[\widetilde{M},M\right]_\ast\right]_\ast\right]_\ast\right]_\ast
		+\left[\widetilde{M},\left[M,\left[M,\left[M,\widetilde{M}\right]_\ast\right]_\ast\right]_\ast\right]_\ast\right)\\
	&\qquad+\frac{1}{120}\left(
		\left[\widetilde{M},\left[M,\left[\widetilde{M},\left[M,\widetilde{M}\right]_\ast\right]_\ast\right]_\ast\right]_\ast
		+\left[M,\left[\widetilde{M},\left[M,\left[\widetilde{M},M\right]_\ast\right]_\ast\right]_\ast\right]_\ast\right)
	-\mathcal{O}\left(\left[\quad\right]_\ast^5\right)
\end{aligned}
\ee
A nontrivial check for the formula \eqref{BCHFreeEnerg}
with a simpler situation will be given in the following subsection.

By the formula of genus expansion for $M^{1D}$, \eqref{FEGenExp},
and the formula \eqref{BCHFreeEnerg}, together with the definition of the
Moyal--Weyl product \eqref{MWprod}, it is transparent that
the free energy $M^{\rm TM}_\ast$ has a genus expansion in $\lambda$ of form
\be
	M^{\rm TM}_\ast(t,\widetilde{t},\k)
	=\sum_{g=0}^\infty \lambda^{2g-2}\,{M}^{\rm TM}_{\ast(g)}(t,\widetilde{t},\k)
\ee
Hence, for example, the lowest order in $\lambda$-expansion is given systematically by
\be
\begin{aligned}
	M^{\rm TM}_{\ast(0)}(t,\widetilde{t},\k)
	&=M_{(0)}+\widetilde{M}_{(0)}
	+\k\left\{M_{(0)},\widetilde{M}_{(0)}\right\}_\ast
	\\&\qquad
	+\frac{\k^2}{3}\left(
	\left\{M_{(0)},\left\{M_{(0)},\widetilde{M}_{(0)}\right\}_\ast\right\}_\ast
	+\left\{\widetilde{M}_{(0)},\left\{\widetilde{M}_{(0)},M_{(0)}\right\}_\ast\right\}_\ast
	\right)
	\\&\qquad
	-\frac{\k^3}{3}
	\left\{\widetilde{M}_{(0)},\left\{M_{(0)},\left\{M_{(0)},\widetilde{M}_{(0)}\right\}_\ast\right\}_\ast\right\}_\ast
	-\mathcal{O}\left(\k^4\left\{\quad\right\}_\ast^4\right)
\end{aligned}
\ee
where $\{\, \cdot \, , \, \cdot \, \}_\ast$ stands for the lowest order of the commutator
\be
	\left[f,g\right]_\ast=2\k\lambda^2\left\{f,g\right\}_\ast+\mathcal{O}(\k^2\lambda^4)
\ee
being regarded as the Poisson bracket
\be
	\left\{f,g\right\}_\ast
	={\frac{\p f}{\p t_0}}{\frac{\p g}{\p \widetilde{t}_0}}
	-{\frac{\p f}{\p \widetilde{t}_0}}{\frac{\p g}{\p t_0}}
\ee
A generalization to any order in $\lambda$-expansion is readily applied.

\subsection{A simpler check for the BCH formula}

As argued above, the free energy of the two-matrix model extension
can be expressed in terms of a pair of the free energies of the one-matrix model
by exploiting the Moyal--Weyl product and its associated
Baker--Campbell--Hausdorff formula.
This subsection provides a simpler but nontrivial check for the formula
\eqref{WeylExp}, as well as \eqref{BCHFreeEnerg},
with a simpler situation
where all the couplings, except $t_0$ and $\widetilde{t}_0$, are turned off.
Under this setup, the partition function reduces to the Gaussian integral,
which can be performed exactly resulting in a form
\bea
	{Z}^{\rm TM}(t_0;\widetilde{t}_0;\k)
	&=\frac{1}{{2 \pi}\lambda^2} 
	\int^\infty_{-\infty} \int^\infty_{-\infty} 
	e^{\frac{1}{\lambda^{2}} 
	\left(-\frac{x^{2}}{2}+ t_{0}x \right)
	+\frac{1}{\lambda^{2}}
	\left(-\frac{y^{2}}{2}+ \widetilde{t}_{0}y \right)
	+\k\frac{xy}{\lambda^2}} dxdy
	\\
	&=\frac{1}{\sqrt{1-\k^2}}\
	\exp\left(
	{\frac{t_0^2+\widetilde{t}_0^2+2\k t_0\widetilde{t}_0}{2\lambda^2(1-\k^2)}}
	\right)
\ena
Expanding the resultant in a Taylor series in $\t,\ \s,\ \k$,
neglecting $\mathcal{O}(\t^4,\s^4,\k^4)$, one finds
\be
\label{ZTaylor}
\begin{aligned}
	{Z}^{\rm TM}(t_0;\widetilde{t}_0;\k)
	&=
	1+{{\s^2}\over{2\r^2}}+{{\t^2}\over{2\r^2}}+{{\t^2\s^2}\over{4\r^4}}
	+\k\left(
	{{\t\s}\over{\r^2}}+{{\t\s^3}\over{2\r^4}}+{{\t^3\s}\over{2\r^4}}
	+{{\t^3\s^3}\over{4\r^6}}
	\right)
	\\&\qquad
	+\k^2\left(
	{{1}\over{2}}+{{3\s^2}\over{4\r^2}}+{{3\t^2}\over{4\r^2}}
	+{{9\t^2\s^2}\over{8\r^4}}
 	\right)
	+\k^3\left(
	{{3\t\s}\over{2\r^2}}+{{5\t\s^3}\over{4\r^4}}
	+{{5\t^3\s}\over{4\r^4}}+{{25\t^3\s^3}\over{24\r^6}}
	\right)
	+\dots
\end{aligned}
\ee

While, on the other, the formula \eqref{WeylExp} reads
\be
\label{WeylExp00}
	{Z}^{\rm TM}(t_0;\widetilde{t}_0;\k)
	={\rm Exp}_\ast\left(M^{\rm TM}_\ast(t_0;\widetilde{t}_0;\k)\right)
	=\sum_{n=0}^\infty\frac{1}{n!}
	\left(M^{\rm TM}_\ast(t_0;\widetilde{t}_0;\k)\right)^{\ast (n)}
\ee
where $M^{\rm TM}_\ast(t_0;\widetilde{t}_0;\k)$ for the situation under consideration
is now given by the formula \eqref{BCHFreeEnerg} with
\be
	M=\frac{t_0^2}{2\lambda^2}
	\qquad\qquad
	\widetilde{M}=\frac{\widetilde{t}_0^2}{2\lambda^2}
\ee
then
\be
	\left[ M,\widetilde{M}\right]_\ast
	=2\frac{\kappa}{\lambda^2}t_0\widetilde{t}_0
	\qquad\quad
	\left[ M,\left[ M,\widetilde{M}\right]_\ast \right]_\ast
	=4\frac{\kappa^2}{\lambda^2}t_0^2
	\qquad\quad
	\left[ \widetilde{M},\left[ \widetilde{M},{M}\right]_\ast \right]_\ast
	=4\frac{\kappa^2}{\lambda^2}\widetilde{t}_0^2
\ee
and
\be
	\left[\widetilde{M},\left[M,\left[M,\widetilde{M}\right]_\ast\right]_\ast\right]_\ast
	=-16\frac{\k^3}{\r^2}\t\s
\ee
Henceforth, neglecting $\mathcal{O}(\t^4,\s^4,\k^4)$, one finds
\bea
	\left(M^{\rm TM}_\ast(t_0;\widetilde{t}_0;\k)\right)^{\ast (0)}
	&=1
	\\
	\left(M^{\rm TM}_\ast(t_0;\widetilde{t}_0;\k)\right)^{\ast (1)}
	&=M^{\rm TM}_\ast(t_0;\widetilde{t}_0;\k)
	\\
	&={{\t^2}\over{2\r^2}}+{{\s^2}\over{2\r^2}}
	+\k\,{{\t\s}\over{\r^2}}
	+\k^2\left({{\t^2}\over{3\r^2}}+{{\s^2}\over{3\r^2}}\right)
	+\k^3\,{{2\t\s}\over{3\r^2}}+\mathcal{O}(\k^4)
\ena
Referring to an explicit form of the Moyal--Weyl product
\bea
	&\left(M^{\rm TM}_\ast(t_0;\widetilde{t}_0;\k)\right)^{\ast (2)}
	=M^{\rm TM}_\ast \ast \ M^{\rm TM}_\ast
	\\
	&=\left(M^{\rm TM}_\ast\right)^2
	+\left(\k\r^2\right)\left(
	\frac{\p M^{\rm TM}_\ast}{\p\t}\frac{\p M^{\rm TM}_\ast}{\p\s}
	-\frac{\p M^{\rm TM}_\ast}{\p\s}\frac{\p M^{\rm TM}_\ast}{\p\t}
	\right)
	\nn
	&\qquad
	+\frac{\left(\k\r^2\right)^2}{2!}\left(
	\frac{\p^2 M^{\rm TM}_\ast}{\p\t^2}\frac{\p^2 M^{\rm TM}_\ast}{\p\s^2}
	-2\frac{\p^2 M^{\rm TM}_\ast}{\p\t\p\s}\frac{\p^2 M^{\rm TM}_\ast}{\p\t\p\s}
	+\frac{\p^2 M^{\rm TM}_\ast}{\p\s^2}\frac{\p^2 M^{\rm TM}_\ast}{\p\t^2}
	\right)
	\nn
	&\qquad
	+\frac{\left(\k\r^2\right)^3}{3!}\left(
	\frac{\p^3 M^{\rm TM}_\ast}{\p\t^3}\frac{\p^3 M^{\rm TM}_\ast}{\p\s^3}
	-3\frac{\p^3 M^{\rm TM}_\ast}{\p\t^2\p\s}\frac{\p^3 M^{\rm TM}_\ast}{\p\t\p\s^2}
	+3\frac{\p^3 M^{\rm TM}_\ast}{\p\t\p\s^2}\frac{\p^3 M^{\rm TM}_\ast}{\p\t^2\p\s}
	-\frac{\p^3 M^{\rm TM}_\ast}{\p\s^3}\frac{\p^3 M^{\rm TM}_\ast}{\p\t^3}
	\right)
	\nn
	&\qquad+\mathcal{O}(\k^4)
\ena
a direct evaluation shows
\be
	\left(M^{\rm TM}_\ast(t_0;\widetilde{t}_0;\k)\right)^{\ast (2)}
	={{\t^2\s^2}\over{4\r^4}}
	+\k\left(
	{{\t\s^3}\over{2\r^4}}+{{\t^3\s}\over{2\r^4}}
	\right)
	+\k^2\left(
	{{1}\over{2}}+{{5\t^2\s^2}\over{6\r^4}}
	\right)
	+\k^3\left(
	{{2\t\s^3}\over{3\r^4}}+{{2\t^3\s}\over{3\r^4}}
	\right)+\mathcal{O}(\t^4,\s^4,\k^4)
\ee
and, in a similar manner, up to terms concerned with the truncation at
$\mathcal{O}(\t^4,\s^4,\k^4)$,
\bea
	\left(M^{\rm TM}_\ast(t_0;\widetilde{t}_0;\k)\right)^{\ast (3)}
	&=\k \, {{\t^3\s^3}\over{4\r^6}} 
	+\k^2\left(
	{{5\t^2}\over{12\r^2}}+{{5\s^2}\over{12\r^2}}
	\right)
	+\k^3\left(
	{{5\t\s}\over{6\r^2}}+{{2\t^3\s^3}\over{3\r^6}}
	\right)+\mathcal{O}(\t^4,\s^4,\k^4)
	\\
	\left(M^{\rm TM}_\ast(t_0;\widetilde{t}_0;\k)\right)^{\ast (4)}
	&=\k^2\, {{7\t^2\s^2}\over{24\r^4}}
	+\k^3\left(
	{{7\t\s^3}\over{12\r^4}}+{{7\t^3\s}\over{12\r^4}}
	\right)+\mathcal{O}(\t^4,\s^4,\k^4)
	\\
	\left(M^{\rm TM}_\ast(t_0;\widetilde{t}_0;\k)\right)^{\ast (5)}
	&=\k^3\, {{3\t^3\s^3}\over{8\r^6}}+\mathcal{O}(\t^4,\s^4,\k^4)
\ena
Combining them altogether with an appropriate factorial given by \eqref{WeylExp00},
the series expansion \eqref{ZTaylor} is reproduced.

\subsection{Extension to generic potentials}

A generalization of the interaction term
\be
	\k\, xy
	\quad\longrightarrow\quad
	\sum_{m,n\geq0}
	\k_{m,n}\frac{x^{m+1}y^{n+1}}{(m+1)!(n+1)!}
\ee
straightforwardly defines an extension of the partition function
\be
	{Z}^{\rm TM}_{\star}(t,\widetilde{t},\k)
	=\frac{1}{{2 \pi}\lambda^2} 
	\int^\infty_{-\infty} \int^\infty_{-\infty} 
	e^{\frac{1}{\lambda^{2}} S(x)+\frac{1}{\lambda^{2}} \widetilde{S}(y)
	+\frac{1}{\lambda^2}\sum_{m,n\geq0}
	\k_{m,n}\frac{x^{m+1}y^{n+1}}{(m+1)!(n+1)!}} dxdy
\ee
Obviously, following the argument in the preceding subsections, 
the partition function is given by
\be
	{Z}^{\rm TM}_{\star}(t,\widetilde{t},\k)
	=Z^{1 D}(t)\star Z^{1 D}(\widetilde{t})
\ee
with
\be
	f(t,\widetilde{t})\star g(t,\widetilde{t})
	=f(t,\widetilde{t})\,e^{\frac{i}{2}
	\sum_{m,n\geq0}\theta_{m,n}
	\left(
	\overset{\leftarrow}{\frac{\p}{\p t_m}}
	\overset{\rightarrow}{\frac{\p}{\p \widetilde{t}_n}}
	-\overset{\leftarrow}{\frac{\p}{\p \widetilde{t}_n}}
	\overset{\rightarrow}{\frac{\p}{\p t_m}}\right)}
	g(t,\widetilde{t})
\ee
and
\be
	\theta_{m,n}=-2i\k_{m,n}\lambda^2
\ee
It shall be apparent that the partition function has a property
\be
	\frac{1}{\lambda^2} \,\frac{\p {Z}^{\rm TM}_\star}{\p\k_{m,n}}(t,\widetilde{t},\k)
	=\frac{\p Z^{1 D}}{\p t_m}(t)\star \frac{\p Z^{1 D}}{\p \widetilde{t}_n}(\widetilde{t})
\ee
and, owing to the fact that $ Z^{1 D}$ obeys the Burgers hierarchy,
\be
	\frac{1}{\lambda^2} \,\frac{\p {Z}^{\rm TM}_\star}{\p\k_{m,n}}(t,\widetilde{t},\k)
	=\frac{\lambda^{2(m+n)}}{(m+1)!(n+1)!}
	\frac{\p^{m+1} Z^{1 D}}{\p t_0^{m+1}}(t)\star 
	\frac{\p^{n+1} Z^{1 D}}{\p \widetilde{t}_0^{n+1}}(\widetilde{t})
\ee
identifying $\k_{0,-1}\equiv t_0$ and $\k_{-1,0}\equiv \widetilde{t}_0$
\be
\label{BurgHierTMMPartFunct}
	\frac{(m+1)!(n+1)!}{\lambda^{2(m+n)}}\,
	\frac{1}{\lambda^2} \,\frac{\p}{\p\k_{m,n}}{Z}^{\rm TM}_\star
	=
	\frac{\p^{m+1}}{\p \k_{0,-1}^{m+1}}
	\frac{\p^{n+1}}{\p \k_{-1,0}^{n+1}} {Z}^{\rm TM}_\star
\ee
which is to be thought of as an extended version of the relation \eqref{BZ}
that is equivalent to the Burgers hierarchy.
This observation shows that the two-matrix analogue of one-dimensional topological
gravity contains Burgers hierarchy partly, but has some extended structure.
The formula for the ``free energy'' $M^{\rm TM}_\star(t,\widetilde{t},\k)$ defined 
through the relation
\be
	{Z}^{\rm TM}_\star(t,\widetilde{t},\k)
	={\rm Exp}_\star \left( M^{\rm TM}_\star(t,\widetilde{t},\k) \right)
\ee
shall be given by \eqref{BCHFreeEnerg} with replacing $\ast \to \star$.

\section{Noncommutative Burgers hierarchy}

\subsection{Noncommutative Burgers equation and hierarchy structure}

This subsection focuses on a hierarchy structure responsible for the noncommutative
description. First, for simplicity, the argument is restricted to 
the relation concerning $t_1=\k_{1,-1}$ and $t_0=\k_{0,-1}$ only,
which leads to the noncommutative Burgers equation.
Based on the recurrence relation along with a family of $\{t_n=\k_{n,-1}\}$, 
it is argued that the relation will be identified with noncommutative Burgers hierarchy
under specifications of the objects in a relevant manner.

Before going to the hierarchy structure,
some preliminary formulae shall be noted.
The differential of the exponential map obeys the following formula
\be
	\frac{\p}{\p \k_{m,n}}{Z}^{\rm TM}_\star
	=\frac{\p}{\p \k_{m,n}}{\rm Exp}_{\star}\left(M^{\rm TM}_\star\right)
	=\frac{D M^{\rm TM}_\star}{D\k_{m,n}}
	\star\, {Z}^{\rm TM}_\star
\ee
where
\be
\label{DiffDD}
	\frac{D M^{\rm TM}_\star}{D\k_{m,n}}
	=
	\sum_{n=0}^\infty\frac{1}{n!}\ {\rm Ad}_\star^{(n)}
	\left(\frac{\p M^{\rm TM}_\star}{\p\k_{m,n}}\right)
\ee
with ${\rm Ad}_\star^{(n)}$ being recursively defined by
\be
	{\rm Ad}_\star^{(0)}\left(\frac{\p M^{\rm TM}_\star}{\p\k_{m,n}}\right)
	=\frac{\p M^{\rm TM}_\star}{\p\k_{m,n}}
	\qquad\quad
	{\rm Ad}_\star^{(n)}\left(\frac{\p M^{\rm TM}_\star}{\p\k_{m,n}}\right)
	=\left[M^{\rm TM}_\star, \
	{\rm Ad}_\star^{(n-1)}\left(\frac{\p M^{\rm TM}_\star}{\p\k_{m,n}}\right)
	\right]_{\star}
\ee
The commutativity for the order of the differentials of the partition function
\be
\label{CommDiff}
	\frac{\p}{\p \k_{m,n}}\frac{\p}{\p \k_{m',n'}}{Z}^{\rm TM}_\star
	=
	\frac{\p}{\p \k_{m',n'}}\frac{\p}{\p \k_{m,n}}{Z}^{\rm TM}_\star
\ee
gives rise to a relation
\be
\label{FlipDiff}
	\frac{\p}{\p \k_{m,n}}\frac{D M^{\rm TM}_\star}{D\k_{m',n'}}
	=
	\frac{\p}{\p \k_{m',n'}}\frac{D M^{\rm TM}_\star}{D\k_{m,n}}
	-\left[\frac{D M^{\rm TM}_\star}{D\k_{m',n'}},
	\frac{D M^{\rm TM}_\star}{D\k_{m,n}}\right]_{\star}
\ee

Focusing on the relation \eqref{BurgHierTMMPartFunct} with concerning
only $t_1=\k_{1,-1}$ and $t_0=\k_{0,-1}$
\be
	\frac{2!}{\lambda^2} \,\frac{\p}{\p t_1}{Z}^{\rm TM}_\star
	=
	\frac{\p^{2}}{\p t_0^{2}}{Z}^{\rm TM}_\star
\ee
yielding an identity
\be
\label{DmDt1}
	\frac{2!}{\lambda^2} \,\frac{D M^{\rm TM}_\star}{D t_1}
	=\frac{\p}{\p t_0}\frac{D M^{\rm TM}_\star}{D t_0}
	+\frac{D M^{\rm TM}_\star}{D t_0}\star\frac{D M^{\rm TM}_\star}{D t_0}
\ee
Taking a further derivative of \eqref{DmDt1} with respect to $t_0$ and
applying the formula \eqref{FlipDiff}, one finds
\bea
	\frac{2!}{\lambda^2} \,\frac{\p}{\p t_0}\frac{D M^{\rm TM}_\star}{D t_1}
	&=\frac{2!}{\lambda^2}\left(
	\frac{\p}{\p t_1}\frac{D M^{\rm TM}_\star}{D t_0}
	-
	\left[
	\frac{D M^{\rm TM}_\star}{D t_1},\frac{D M^{\rm TM}_\star}{D t_0}
	\right]_\star
	\right)
	\\
	&=\frac{\p^2}{\p t_0^2}\frac{D M^{\rm TM}_\star}{D t_0}
	+\left(\frac{\p}{\p t_0}\frac{D M^{\rm TM}_\star}{D t_0}\right)
		\star\frac{D M^{\rm TM}_\star}{D t_0}
	+\frac{D M^{\rm TM}_\star}{D t_0}
		\star\frac{\p}{\p t_0}\frac{D M^{\rm TM}_\star}{D t_0}
\ena
Inside the right hand side of the first line, 
substituting the relation \eqref{DmDt1} again, results in
\be
\label{NCBurgEq}
	\frac{2!}{\lambda^2}\frac{\p U_\star}{\p t_1}
	=
	\frac{\p^2 U_\star}{\p t_0^2}
	+2\frac{\p U_\star}{\p t_0}\star U_\star
	-\left[U_\star\,,\,\left(U_\star\star U_\star\right)\,\right]_\star
\ee
with introducing
\be
\label{DefOfU}
	U_\star
	=\frac{D M^{\rm TM}_\star}{D t_0}
	=:\frac{D}{Dt_0}{\rm Log}_{\star}\left({Z}^{\rm TM}_\star\right)
\ee
which can be thought of as a noncommutative version of
the Cole--Hopf transformation \eqref{CHtransform}.
The partial differential equation \eqref{NCBurgEq} is to be interpreted as
a noncommutative extension of the Burgers equation.
Indeed, under the commutative limit where $\k\to0$, the equation reduces to
the ordinary Burgers equation \eqref{BurgersEq}.

The identity \eqref{DmDt1} is readily extended to $t_n$:
\be
\label{recurre:full:NCBurHier}
	\frac{(n+1)!}{\lambda^{2n}} \,\frac{D M^{\rm TM}_\star}{D t_n}
	=\left(\frac{\p}{\p t_0}+\frac{D M^{\rm TM}_\star}{D t_0}\right)^{\star(n)}
	\star\frac{D M^{\rm TM}_\star}{D t_0}
\ee
which is a noncommutative counterpart of 
the recurrence relation \eqref{recurre:full:BurHier}.
In a similar manner to the case with $n=1$,
taking a once more derivative of \eqref{recurre:full:NCBurHier} with respect to $t_0$ and
applying the formula \eqref{FlipDiff}, one finds
\be
\label{NCBurgHierarchy}
	\frac{(n+1)!}{\lambda^{2n}}\,\frac{\p U_\star}{\p t_n}
	=\frac{\p}{\p t_0}\left(
	\left(\frac{\p}{\p t_0}+U_\star\right)^{\star(n)}\,\star\, U_\star\right)
	-
	\left[U_\star\,,\,
	\left(\frac{\p}{\p t_0}+U_\star\right)^{\star(n)}\,\star\, U_\star
	\right]_\star
\ee
which is to be identified with a noncommutative extension of 
the Burgers hierarchy \eqref{BurgHirach}.
For example, the case with $n=2$ gives rise to an equation
\be
\label{NCSTOEq}
\begin{aligned}
	\frac{3!}{\lambda^{4}}\,\frac{\p U_\star}{\p t_2}
	&=\frac{\p^3 U_\star}{\p t_0^3}
	+3\frac{\p}{\p t_0}\left(\frac{\p U_\star}{\p t_0}\star U_\star\right)
	+3\frac{\p U_\star}{\p t_0}\star U_\star^{\star(2)}
	\\
	&\qquad
	+\left[U_\star \,,\,\frac{\p^2 U_\star}{\p t_0^2}\right]_\star
	+\left[U_\star \,,\,\frac{\p U_\star}{\p t_0}\right]_\star\star U_\star
	+U_\star\star \left[\frac{\p U_\star}{\p t_0}\,,\,U_\star \right]_\star
	-\left[U_\star \,,\,U_\star^{\star(3)}\right]_\star
\end{aligned}
\ee
where the second line is of order $\mathcal{O}(\k)$, so that only
the first line survives under the commutative limit reproducing the
 Sharma--Tasso--Olver equation \eqref{STOeq}.

In this regard, one can fairly conclude that the partition function of the
two-matrix model extension of the one-dimensional topological gravity is
a tau-function of a noncommutative extension of the Burgers hierarchy defined by
\eqref{NCBurgHierarchy}. The same structure can be shown for the family
$\{\widetilde{t}_n=\k_{-1,n}\}$ as well in a parallel manner except a replacement of
the fundamental object by
\be
	U_\star=\frac{D M^{\rm TM}_\star}{D t_0}
	\quad\longrightarrow\quad
	\widetilde{U}_\star=\frac{D M^{\rm TM}_\star}{D \widetilde{t}_0}
\ee

\subsection{Relation to noncommutative U(1) gauge theory}

The final observation before closing this section is on the relation concerning
$\t$ and $\s$, between which
the noncommutativity is turned on in a sense of \eqref{NCrelation}.
Noting that the formula \eqref{FlipDiff} can also be applied to
$\t=\k_{0,-1}$ and $\s=\k_{-1,0}$, one finds
\be
\label{FlatConn}
	\frac{\p \widetilde{U}_\star}{\p \t}-\frac{\p U_\star}{\p \s}
	-\left[U_\star\,,\,\widetilde{U}_\star\right]_{\star}=0
\ee
which provides a basic relation governing the way
how the two noncommutative Burgers hierarchies shall be intertwined.

Although its intuitive understanding is still missing,
let it be noted that
the equation \eqref{FlatConn} can be interpreted as the condition for the flat connection
for the noncommutative (NC) U(1) gauge field.
Once taking a viewpoint of NC U(1) gauge theory,
the NC Burgers equation \eqref{NCBurgEq} turns out to be nothing but
the other components of the flatness condition on the NC U(1) gauge field strength
defined by
\be
	F_{ab}=\frac{\p A_b}{\p x^a}-\frac{\p A_a}{\p x^b}-\left[A_a,A_b\right]_\star
\ee
on a space labelled by coordinates $\{x^a\}=(\t,t_1,\s,\widetilde{t}_1)$ under identification
\be
\begin{aligned}
	A_{t_0}&=U_\star
	\ \left(=\frac{D M^{\rm TM}_\star}{D t_0}\right)
	\qquad\qquad
	A_{t_1}=\frac{\p U_\star}{\p \t}+U_\star\star U_\star
	\ \left(=\frac{2!}{\lambda^2} \,\frac{D M^{\rm TM}_\star}{D t_1}\right)
	\\
	A_{\widetilde{t}_0}&=\widetilde{U}_\star
	\ \left(=\frac{D M^{\rm TM}_\star}{D \widetilde{t}_0}\right)
	\qquad\qquad
	A_{\widetilde{t}_1}
	=\frac{\p \widetilde{U}_\star}{\p \s}+\widetilde{U}_\star\star \widetilde{U}_\star
	\ \left(=\frac{2!}{\lambda^2} \,\frac{D M^{\rm TM}_\star}{D \widetilde{t}_1}\right)
\end{aligned}
\ee
Indeed, the relation \eqref{FlatConn} is equivalent to $F_{t_0\widetilde{t}_0}=0$,
the NC Burgers equation \eqref{NCBurgEq} is $F_{t_0t_1}=0$,
and that for $\widetilde{U}_\star$ is $F_{\widetilde{t}_0\widetilde{t}_1}=0$.
Additionally, including the other nontrivial components
$F_{t_0\widetilde{t}_1},\ F_{t_1\widetilde{t}_0},\ F_{t_1\widetilde{t}_1}$,
its flatness $F_{ab}=0$ turns out to be a consequence of the commutativity
of the differential \eqref{CommDiff} and thus \eqref{FlipDiff}.

Therefore, the system under consideration can also be
interpreted in terms of the four-dimensional NC U(1) gauge theory on 
$(\t,t_1,\s,\widetilde{t}_1)$ with noncommutativity turned on between
$\t$ and $\s$ (and between $t_1$ and $\widetilde{t}_1$).
The appearance of the NC Burgers equation from the  (anti-)self-dual
(NC) Yang--Mills equation is an example of what the Ward conjecture claims \cite{Ward}.
A similar consideration can be found in \cite{Hamanaka},
where the identification of the gauge fields is different,
giving rise to a different type of NC Burgers equation.
It is noted that
the argument so far on a set $(\t,t_1,\s,\widetilde{t}_1)$
can as well be applied to any set $(t_m,t_n,\widetilde{t}_m,\widetilde{t}_n)$.

Both i) the relationship of $U_\star$ and $\widetilde{U}_\star$ to NC U(1) gauge fields,
and ii) the multiplicity, or in other words, nonlinearity, in the construction of
NC gauge field configurations
from the commutative objects,
e.g. \eqref{BCHFreeEnerg} and \eqref{DefOfU} (that is, \eqref{DiffDD}),
are quite suggestive of the Seiberg--Witten map \cite{Seiberg},
which relates a description by gauge field 
on an ordinary space but with some background field,
to another description by NC U(1) gauge field.
Revealing the relation between the NC Burgers hierarchy and the Seiberg--Witten map
would be an interesting issue to explore, though, it is
beyond the scope of this article and left for another study.

\section{Conclusion and discussion}

A system extended to two-matrix analogue of one-dimensional gravity
has been investigated and it has shown that its free energy consists of
a pair of one-dimensional topological gravity intertwined by the Moyal--Weyl product.
With help of the Baker--Campbell-Hausdorff formula,
the formula for the free energy is obtained and 
it has shown to allow a form of $\lambda$-expansion.
The partition function of the extended model is subject to a hierarchy structure partly 
containing two sets of Burgers hierarchies.
The structure has been shown to be carried over in the description
of noncommutative language, where the noncommutative
counterpart of the Cole--Hopf transformation plays a role
to figure out a noncommutative extension of the Burgers hierarchy.
In this sense, the partition function of what was considered in this article
has been shown to be a tau-function of
the noncommutative Burgers hierarchy.
Under identification of the system with noncommutative U(1) gauge theory,
the noncommutative Burgers equation has been shown to be equivalent to
the condition for the flat connection, which 
is a consequence of the commutativity in differential with respect to
the coupling constants.

An interesting direction for further investigation
would be exploring the integrable structure
of the extended model, although it has been shown at least containing two sets of
noncommutative Burgers hierarchies. 
The corresponding Lax formulation, for example, would be one of possible things to be sought.
A clue might be found in noncommutative 
Burgers equation \cite{Hamanaka}, where noncommutativity is introduced in either
$(t_0,t_1)$ or $(\widetilde{t}_0,\widetilde{t}_1)$, although, in this sense,
it is not the case directly applicable to what was considered in this article.

A crucial difference of what has been found in this article
from two-matrix models is that the interaction
between two one-matrix models can be factored by themselves.
This factorization has played a key role in representing the partition function
as the Moyal--Weyl product. Thus the results obtained in this article
is not straightforwardly applied to those of matrix models.
An investigation along this line, together with the above mentioned direction,
shall be worth exploring.

\subsection*{Another extension of one-dimensional topological gravity}

The open KdV hierarchy is an extension of the KdV hierarchy appearing 
in two-dimensional gravity on closed surfaces to that on surfaces with boundaries 
\cite{PST_14,B_15,BT_17,A1410,A1412},
for recent developments, see \cite{DW}.
An additional equation ($s$-flow equation, and its Laplace transformed version) 
compatible with the open KdV hierarchy was advocated in \cite{B_16} and
it played an important role in providing an explicit solution to the open KdV hierarchy
\cite{Bawane,Muraki,Alexandrov}, where its relation to matrix model description was 
investigated also.
However, when it comes to implications of the open KdV hierarchy,
together with the $s$-flow equation,
still has it not completely been demystified yet,
in a sense of lacking a procedure how to include multiple boundary parameters
that accounts for boundary conditions,
leading to some redundant description
of its matrix model formulation due to combined with bulk coupling constants \cite{IMR}.
Before closing this article, let us here make some comments
on another extension of one-dimensional topological gravity to its open analogue
in view of a rather simpler setting considered throughout this article.

\subsubsection*{Resolvent as generating function}

The loop operator
\be
	W_k(z_1,\dots,z_k;t)
	=\ll \frac{1}{z_1-x}\dots \frac{1}{z_k-x}\rr
	=\frac{1}{\sqrt{2 \pi} \lambda} 
	\int \frac{1}{z_1-x}\dots \frac{1}{z_k-x}\ e^{\frac{1}{\lambda^{2}} S(x)} d x
\ee
has its own importance as generating function:
an expansion in $z_i$ gives
\be
	W_k(z_1,\dots,z_k;t)=
	\sum_{n_1,\dots,n_k=0}^\infty 
	\frac{\ll x^{n_1+\dots+n_k}\rr }{{z_1}^{n_1+1}\dots{z_k}^{n_k+1}}
\ee
in particular $k=1$ provides a generating function of expectation values of monomial in $x$ via
\be
\label{resolvent}
	W_1(z;t)=\ll \frac{1}{z-x}\rr=\sum_{n=0}^\infty \frac{\ll x^n\rr }{z^{n+1}}
\ee
Thus applying Cauchy's residue theorem, the expectation value can be extracted by
\be
	\ll x^n\rr=\oint \frac{dz}{2\pi i} \, z^n\,W_1(z;t)
\ee

\subsubsection*{Open Virasoro constraints}

Rewriting the relation \eqref{resolvent}, 
$W_1(z;t)$ can be regarded as the Laplace transform
\be
	W_1(z;t)
	=\ll \int_0^\infty  dl\, e^{-l(z-x)} \rr
	= \int_0^\infty  dl\,e^{-lz}\,Z^o(l;t)
\ee
with
\be
	Z^o(l;t)
	=\ll e^{lx} \rr
	=\frac{1}{\sqrt{2 \pi} \lambda} 
		\int_{-\infty}^\infty \,e^{lx+\frac{1}{\lambda^{2}} S(x)} d x
\ee
and then
\be
	Z^o(l;t)= Z^{1 D}(t_0+l\lambda^2;t_{i\geq1})
	=e^{l\lambda^2\frac{\p}{\p t_0}}Z^{1 D}(t)
\ee
which might be regarded as one-dimensional version of
open/closed dualities of two-dimensional topological gravity \cite{DJMW,J},
and for more recent information \cite{AT}.
Recalling that $Z^{1 D}(t)$ is subject to the Virasoro constraints \eqref{ViraConZ},
$Z^o(l;t)$ shall also be constrained:
\be
	0=e^{l\lambda^2\frac{\p}{\p t_0}}\left(L_mZ^{1 D}(t)\right)
	=\left(e^{l\lambda^2\frac{\p}{\p t_0}}L_me^{-l\lambda^2\frac{\p}{\p t_0}}\right)Z^o(l;t)
	=:\widetilde{L}_mZ^o(l;t)
\ee
where
\be
	\widetilde{L}_{m}
	=e^{l\lambda^2\frac{\p}{\p t_0}}L_me^{-l\lambda^2\frac{\p}{\p t_0}}
	=
	\begin{cases}
	\ L_{-1}+l & (m=-1)
	\\
	\ L_{0}+l\lambda^2\frac{\p}{\p t_0} & (m=0)
	\\
	\ L_{m}+l\lambda^2(m+1)!\frac{\p}{\p t_m} & (m\geq1)
	\end{cases}
\ee
satisfying the relation
\be
	[\widetilde{L}_{m},\widetilde{L}_{n}]=(m-n)\widetilde{L}_{m+n}
\ee
Then on $W_1(z;t)$ these generators can be represented by
\be
	0
	=\int_0^\infty  dl\,e^{-lz}\,\widetilde{L}_mZ^o(l;t)
	=
	\begin{cases}
	\ \left(L_{-1}-\frac{\p}{\p z}\right)W_1(z;t)
	& (m=-1)\\
	\ \left(L_{0}-\lambda^2\frac{\p}{\p t_0}\frac{\p}{\p z}\right)W_1(z;t)
	& (m=0)\\
	\ \left(L_{m}-(m+1)!\lambda^2\frac{\p}{\p t_m}\frac{\p}{\p z}\right)W_1(z;t)
	& (m\geq1)
	\end{cases}
\ee
It is noted, using the relation \eqref{BZ}, that
\be
	\frac{\p}{\p t_m}W_1(z;t)
	= \frac{\lambda^{2m}}{(m+1)!}\frac{\p^{m+1}}{\p t_0^{m+1}}W_1(z;t)
\ee
then for $m\geq1$
\be
	L^o_{m}W_1(z;t):=
	\left\{L_{m}+
	\left(\lambda^2\frac{\p}{\p t_0}\right)^{m+1}
	\left(-\frac{\p}{\p z}\right)\right\}W_1(z;t)=0
\ee

The generating function $W_1(z;t)$ obeys also another infinite set of identities
\be
\label{OpenVir2}
	\overline{L}_m^o W_1(z;t)=0
\ee
with
\be
	\overline{L}_m^o=\overline{L}_m
	+\left(-\frac{\p}{\p z}\right)^{m+1}\left(\lambda^2\frac{\p}{\p t_0}\right) 
	\qquad\quad
	\overline{L}_m=\left(-\frac{\p}{\p z}\right)^{m+1}\circ(-z)
\ee
which satisfy the (part of) Virasoro algebra
\be
	[\overline{L}_m^o,\overline{L}_n^o]=(m-n)\overline{L}_{m+n}^o
	\qquad\quad
	[\overline{L}_m,\overline{L}_n]=(m-n)\overline{L}_{m+n}
\ee
These identities \eqref{OpenVir2} follow from invariance under
reparametrization $l\to l+\varepsilon_m l^{m+1}\ (m\geq0)$:
\bea
	W_1(z;t)
	&=\ll \int_0^\infty dl
	\left(1+(m+1)\varepsilon_m l^{m}-\varepsilon_m l^{m+1}(z-x)\right)
	\, e^{-l(z-x)} \rr
	\nn
	&=W_1(z;t)
	+\varepsilon_m\left\{(m+1)\left(-\frac{\p}{\p z}\right)^m-
	\left(z-\lambda^2\frac{\p}{\p t_0}\right)\left(-\frac{\p}{\p z}\right)^{m+1}\right\}
	W_1(z;t)
	\nn
	&=W_1(z;t)
	+\varepsilon_m\,\overline{L}_m^o\,W_1(z;t)
\ena
Although we have derived two sets of (part of) Virasoro generators, $\{L^o_{m}\}$
and $\{\overline{L}^o_{m}\}$,
their implications, in particular their associated hierarchy structure
(if it exists), are still missing. 
Further investigation on this aspect shall also be carried out.

\subsection*{Acknowledgements}

This research was supported by an appointment to the YST Program at the APCTP 
through the Science and Technology Promotion Fund and Lottery Fund of the Korean Government.
This was also supported by the Korean Local Governments - Gyeongsangbuk-do Province 
and Pohang City.

\appendix
\section{Virasoro constraints \label{AppA}}

The constraints can be derived by performing an infinitesimal change of variable $x\to x+\varepsilon_mx^{m+1}\ (m\geq 1)$
\be
\begin{aligned}
	Z^{1 D}(t)
	&=
	\frac{1}{\sqrt{2 \pi} \lambda} \int 
	e^{-\frac{(x+\varepsilon_mx^{m+1})^2}{2\lambda^{2}}
		+\sum_{n=0}^{\infty} \frac{t_{n}}{\lambda^2} \frac{(x+\varepsilon_mx^{m+1})^{n+1}}{(n+1) !}} 
		d(x+\varepsilon_mx^{m+1})
	\\
	&=Z^{1 D}(t)
	+\frac{\varepsilon_m}{\lambda^2}
	\frac{1}{\sqrt{2 \pi} \lambda} \int 
	e^{-\frac{x^2}{2\lambda^{2}}+\sum_{n=0}^{\infty} \frac{t_{n}}{\lambda^2} \frac{x^{n+1}}{(n+1) !}} 
	\left(-x^{m+2}+\sum_{n=0}^{\infty} t_{n}\frac{x^{m+n+1}}{n !}+\lambda^2 (m+1)x^m\right)
	d x
	+\mathcal{O}(\varepsilon_m^2)
	\notag
\end{aligned}
\ee
noting that
\be
	\frac{\p}{\p t_n}Z^{1 D}(t)
	=
	\frac{1}{\sqrt{2 \pi} \lambda} \int 
	e^{-\frac{x^2}{2\lambda^{2}}+\sum_{n=0}^{\infty} \frac{t_{n}}{\lambda^2} \frac{x^{n+1}}{(n+1) !}} 
	\left(\frac{1}{\lambda^2}\frac{x^{n+1}}{(n+1)!}\right)
	d x
\ee
then one finds an identity
\be
	0=
	\varepsilon_n
	\left(
	-(m+2)!\frac{\p}{\p t_{m+1}}+\sum_{n=0}^{\infty} t_{n}\frac{(m+n+1)!}{n !}\frac{\p}{\p t_{m+n}}
	+\lambda^2 (m+1)!\frac{\p}{\p t_{m-1}}
	\right)
	\frac{1}{\sqrt{2 \pi} \lambda} \int 
	e^{-\frac{x^2}{2\lambda^{2}}+\sum_{n=0}^{\infty} \frac{t_{n}}{\lambda^2} \frac{x^{n+1}}{(n+1) !}} 
	d x
	\notag
\ee
thus
\be
	L_{m} \, Z^{1 D}(t)=0
\ee
And $x\to x+\varepsilon_{0}x$ (dilation):
\be
\begin{aligned}
	Z^{1 D}(t)
	&=
	\frac{1}{\sqrt{2 \pi} \lambda} \int 
	e^{-\frac{(x+\varepsilon_{0}x)^2}{2\lambda^{2}}
		+\sum_{n=0}^{\infty} \frac{t_{n}}{\lambda^2} \frac{(x+\varepsilon_{0}x)^{n+1}}{(n+1) !}} 
		(1+\varepsilon_{0})dx
	\\
	&=Z^{1 D}[t]
	+\varepsilon_{0}
	\frac{1}{\sqrt{2 \pi} \lambda} \int 
	e^{-\frac{x^2}{2\lambda^{2}}+\sum_{n=0}^{\infty} \frac{t_{n}}{\lambda^2} \frac{x^{n+1}}{(n+1) !}} 
	\left(-\frac{x^2}{\lambda^2}
		+\sum_{n=0}^{\infty} t_{n}\frac{1}{\lambda^2}\frac{x^{n+1}}{n !}+1\right)
	d x
	+\mathcal{O}(\varepsilon_0^2)
\end{aligned}
	\notag
\ee
thus
\be
	L_{0} \, Z^{1 D}(t)=0
\ee
Finally, $x\to x+\varepsilon_{-1}$ (translation):
\be
\begin{aligned}
	Z^{1 D}(t)
	&=
	\frac{1}{\sqrt{2 \pi} \lambda} \int 
	e^{-\frac{(x+\varepsilon_{-1})^2}{2\lambda^{2}}
		+\sum_{n=0}^{\infty} \frac{t_{n}}{\lambda^2} \frac{(x+\varepsilon_{-1})^{n+1}}{(n+1) !}} 
		dx
	\\
	&=Z^{1 D}[t]
	+\varepsilon_{-1}
	\frac{1}{\sqrt{2 \pi} \lambda} \int 
	e^{-\frac{x^2}{2\lambda^{2}}+\sum_{n=0}^{\infty} \frac{t_{n}}{\lambda^2} \frac{x^{n+1}}{(n+1) !}} 
	\left(-\frac{x}{\lambda^2}+\frac{t_0}{\lambda^2}
		+\sum_{n=1}^{\infty} t_{n}\frac{1}{\lambda^2}\frac{x^{n}}{n !}\right)
	d x
	+\mathcal{O}(\varepsilon_{-1}^2)
\end{aligned}
	\notag
\ee
thus
\be
	L_{-1} \, Z^{1 D}(t)=0
\ee

\section{Change of coupling constants \label{AppB}}

A set of variables is defined by
\be
	I_n=\sum_{m=0}^{\infty} t_{n+m} \frac{x^{m}}{m !}
\ee
which serve an alternative set of couplings $t_k$.
By definition, one finds $x_\i=I_0$ and $I_n=\sum_{m=0}^{\infty} t_{n+m} \frac{I_0^{m}}{m !}$.
Using the variables ($k\geq2$), one finds
\be
	\frac{d^kS}{dx^k}(x)=-\delta_{k,2}+I_{k-1}
\ee
Performing a Taylor expansion of $S(x)$ around the saddle point $x_\i$, one obtains
\bea
	S(x)
	&=S(x_\i)+(x-x_\i)\frac{dS}{dx}(x_\i)+\sum_{k=2}^\infty\frac{(x-x_\i)^k}{k!}\frac{d^kS}{dx^k}(x_\i)\nn
	&=-\frac{1}{2} I_0^{2}+\sum_{n=0}^{\infty} \frac{t_{n}I_0^{n+1}}{(n+1) !}
	+\sum_{k=2}^\infty \left(I_{k-1}-\delta_{k,2}\right)\frac{(x-I_0)^k}{k!}
\ena
From the definition of $I_k$, it follows that
\be
	\sum_{n=0}^{\infty}  \frac{t_{n}I_0^{n}}{(n+1) !}
	=
	\sum_{n=0}^\i\frac{(-1)^n}{(n+1)!}I_nI_0^n
\ee
with a help of formula
\be
	\sum_{k=0}^n  \frac{(-1)^k}{(k+1)!(n-k)!}=\frac{1}{(n+1)!}
\ee
which is shown by 
$0=(1-1)^{n+1}=1-\sum_{k=0}^{n}(-1)^k \binom{n+1}{k+1}$.
Hence the Taylor expansion of $S(x)$ around $x_\i$ reads
\be
	S(x)=
	\sum_{n=0}^\i\frac{(-1)^n}{(n+1)!}\left(I_n+\delta_{n,1}\right)I_0^{n+1}
	+\sum_{n=2}^\infty \left(I_{n-1}-\delta_{n,2}\right)\frac{(x-I_0)^n}{n!}
\ee

One can switch the description in $\{t_k\}$ to that in $\{I_k\}$, and vice versa, as follows.
From
\be
\begin{aligned}
	I_0=\sum_{m=0}^{\infty} t_{m} \frac{I_0^{m}}{m !}
		&=t_0+t_1\frac{I_0}{1!}+t_2\frac{I_0^2}{2!}+t_3\frac{I_0^3}{3!}+t_4\frac{I_0^4}{4!}
			+t_5\frac{I_0^5}{5!}+t_6\frac{I_0^6}{6!}+\dots\\
	I_1I_0=\sum_{m=0}^{\infty} t_{m+1} \frac{I_0^{m+1}}{m !}
		&=\qquad t_1 I_0+t_2\frac{I_0^2}{1!}+t_3\frac{I_0^3}{2!}+t_4\frac{I_0^4}{3!}
			+t_5\frac{I_0^5}{4!}+t_6\frac{I_0^6}{5!}+\dots\\
	I_2I_0^2=\sum_{m=0}^{\infty} t_{m+2} \frac{I_0^{m+2}}{m !}
		&=\qquad\qquad\quad t_2I_0^2+t_3\frac{I_0^3}{1!}+t_4\frac{I_0^4}{2!}
			+t_5\frac{I_0^5}{3!}+t_6\frac{I_0^6}{4!}+\dots\\
	I_3I_0^3=\sum_{m=0}^{\infty} t_{m+3} \frac{I_0^{m+3}}{m !}
		&=\qquad\qquad\qquad\qquad  t_3I_0^3+t_4\frac{I_0^4}{1!}
			+t_5\frac{I_0^5}{2!}+t_6\frac{I_0^6}{3!}+\dots
\end{aligned}
\ee
and so forth, it follows that
\be
	t_n=\sum_{k=0}^\infty\frac{(-1)^kI_0^k}{k!}I_{k+n}
\ee
thus
\be
	\frac{\p t_n}{\p I_m}
	=\frac{(-1)^{m-n}I_0^{m-n}}{(m-n)!}H(m-n)
	-t_{n+1}\delta_{0,m}
\ee
where $H(x)$ is the Heaviside step function
\be
	H(x)=\begin{cases} 1 \qquad (x\geq 0) \\ 0 \qquad (x<0) \end{cases}
\ee
Hence, from the Leibniz rule, the derivative with respect to $I_m$ can be represented as
\be
	\frac{\p}{\p I_m}
	=\sum_{n=0}^\infty\frac{\p t_n}{\p I_m}\frac{\p}{\p t_n}
	=\sum_{n=0}^\infty\left(
	\frac{(-1)^{m-n}I_0^{m-n}}{(m-n)!}H(m-n)-t_{n+1}\delta_{0,m}
	\right)\frac{\p}{\p t_n}
\ee
so that
\be
	\frac{\p}{\p I_0}
	=\frac{\p}{\p t_0}-\sum_{n=0}^\infty t_{n+1}\frac{\p}{\p t_n}
	\qquad\qquad
	\frac{\p}{\p I_m}
	=\sum_{n=0}^m
	\frac{(-1)^{m-n}I_0^{m-n}}{(m-n)!}\frac{\p}{\p t_n}
\ee
Conversely, from
\be
	I_n=\sum_{m=0}^{\infty} t_{n+m} \frac{I_0^{m}}{m !}
\ee
one finds
\be
	\frac{\p I_n}{\p t_k}
	= \frac{I_0^{k-n}}{(k-n) !}H(k-n)
	+I_{n+1}\frac{\p I_0}{\p t_k}
\ee
in particular
\be
	\left(1-\sum_{m=0}^{\infty} t_{m+1} \frac{I_0^{m}}{m !}\right)\frac{\p I_0}{\p t_k}=\frac{I_0^k}{k!}
\ee
so that
\be
	\frac{\p I_0}{\p t_k}=\frac{1}{1-I_1}\frac{I_0^k}{k!}
\ee
and then
\be
	\frac{\p I_n}{\p t_k}
	= \frac{I_0^{k-n}}{(k-n) !}H(k-n)
	+\frac{I_{n+1}}{1-I_1}\frac{I_0^k}{k!}
\ee
Therefore
\be
	\frac{\p}{\p t_k}
	=\sum_{n=0}^\infty\frac{\p I_n}{\p t_k}\frac{\p}{\p I_n}
	=\frac{I_0^k}{k!}
	\left(\frac{1}{1-I_1}\frac{\p}{\p I_0}
	+\sum_{n=1}^\infty\frac{I_{n+1}}{1-I_1}\frac{\p}{\p I_n}
	\right)
	+\sum_{n=1}^k \frac{I_0^{k-n}}{(k-n) !}\frac{\p}{\p I_n}
\ee
Thus, knowing the Jacobian,
one can switch descriptions in $\{t_k\}$ and in $\{I_k\}$.

\section{Consistency checks on recurrence relation \label{AppConsist}}

A consistency check of \eqref{recurre:full}
with the string equation \eqref{stringM} as follows.
Multiplying \eqref{recurre:full} with $t_n$ and summing over $n$, one finds
\bea
	\sum_{n=1}^\infty t_n\times(\text{l.h.s. of \eqref{recurre:full}})
	&=\sum_{n=1}^\infty t_n\frac{n+1}{\lambda^{2}} \frac{\partial M^{1D}}{\partial t_{n}}\nn
	&=\frac{1}{\lambda^2}\left(-t_0\frac{\p M^{1D}}{\p t_{0}}-1\right)
		+\frac{2}{\lambda^2}\frac{\p M^{1D}}{\p t_{1}}
\ena
whereas
\bea
	\sum_{n=1}^\infty t_n\times(\text{r.h.s. of \eqref{recurre:full}})
	&=\frac{\partial M^{1D}}{\partial t_{0}} \sum_{n=1}^\infty t_n\frac{\partial M^{1D}}{\partial t_{n-1}}
	+\sum_{n=1}^\infty t_n\frac{\partial^{2} M^{1D}}{\partial t_{0} \partial t_{n-1}}\nn
	&=\frac{\partial M^{1D}}{\partial t_{0}} \left(-\frac{t_0}{\lambda^2}+\frac{\p M^{1D}}{\p t_{0}}\right)
	+\frac{\p}{\p t_0}\left(-\frac{t_0}{\lambda^2}+\frac{\p M^{1D}}{\p t_{0}}\right)\nn
	&=\frac{1}{\lambda^2}\left(-t_0\frac{\p M^{1D}}{\p t_{0}}-1\right)
	+\frac{\partial M^{1D}}{\partial t_{0}} \frac{\partial M^{1D}}{\partial t_{0}}
	+\frac{\partial^{2} M^{1D}}{\partial t_{0} \partial t_{0}}
\ena
where the first and second relations of \eqref{stringM} are used.
The discrepancy is reduced to $n=1$ case of \eqref{recurre:full}.

A direct evaluation with use of an explicit formula \eqref{freeene1} for $M^{1D}_{(1)}$
also ensures that the relation \eqref{recurre:full} should be valid
for the next to the leading order in $\lambda$-expansion:
\be
	(n+1)\frac{\p M^{1D}_{(1)}}{\p t_n}
	=\sum_{g=0,1}\frac{\partial M^{1D}_{(g)}}{\partial t_{0}} \frac{\partial M^{1D}_{(1-g)}}{\partial t_{n-1}}
		+\frac{\p^2 M^{1D}_{(0)}}{\p {t_0}\p t_{n-1}}
\ee
Crucial relations for its proof are the following
\be
	\frac{\p M^{1D}_{(1)}}{\p t_n}
	=\frac{1}{n}
		\left(\frac{\p M^{1D}_{(0)}}{\p t_0}\frac{\p M^{1D}_{(1)}}{\p t_{n-1}}
		+\frac{1}{2}\frac{\p^2 M^{1D}_{(0)}}{\p t_0\p t_{n-1}}\right)
	=\frac{\p M^{1D}_{(1)}}{\p t_0}\frac{\p M^{1D}_{(0)}}{\p t_{n-1}}
	+\frac12\frac{\p^2 M^{1D}_{(0)}(t)}{\p t_0\p t_{n-1}}
\ee

\section{($G'/G$)-expansion method \label{AppC}} 
 
The ($G'/G$)-expansion method puts an ansatz that 
the solution shall be the form of a (Laurent) polynomial of ($G'/G$):
\be
	u(x,\tau)=u(\xi)=\sum_{n=N_{\rm min}}^{N_{\rm max}} a_n\left(\frac{G'}{G}\right)^n
	\qquad\quad
	G=G(\xi)
\ee
with
\be
	\xi=x-c\tau
	\qquad\quad
	G'=\frac{d G}{d \xi}
\ee
assuming $G$ should solve the following ordinary differential equation
\be
	G''+rG'+sG=0
\ee
All the parameters $a_n,\ r,\ s,\ c$ in the ansatz are to be determined later.
The solutions to the linear differential equation on $G$ are classified in three types
according to the value of its discriminant $D=r^2-4s$:
\be
G(\xi)=
\begin{cases}
	\quad
	e^{-\frac{r\xi}{2}}
	\left\{C_1\sin\left(\frac{\sqrt{4s-r^2}}{2}\xi\right)
		+C_2\cos\left(\frac{\sqrt{4s-r^2}}{2}\xi\right)\right\}
	&\qquad (D<0)
	\\
	\quad
	e^{-\frac{r\xi}{2}}\left(C_1\xi+C_2\right)
	&\qquad (D=0)
	\\
	\quad
	e^{-\frac{r\xi}{2}}
	\left\{C_1\sinh\left(\frac{\sqrt{r^2-4s}}{2}\xi\right)
		+C_2\cosh\left(\frac{\sqrt{r^2-4s}}{2}\xi\right)\right\}
	&\qquad (D>0)
\end{cases}
\ee
A remarkable property and a reason for putting such ansatz
is that a differential with respect to $\xi$ defines an automorphism
on a set of (Laurent) polynomials in $(G'/G)$, because of
\be
	\left[\left(\frac{G'}{G}\right)^n\right]'
	=-n\left(\frac{G'}{G}\right)^{n+1}-rn\left(\frac{G'}{G}\right)^{n}
	-sn\left(\frac{G'}{G}\right)^{n-1}
\ee
where
\be
\frac{G'}{G}=
\begin{cases}
	\quad
	-\frac{1}{2}
	\left\{
	\frac{\left(rC_1+\sqrt{4s-r^2}C_2\right)\sin\left(\frac{\sqrt{4s-r^2}}{2}\xi\right)
		+\left(rC_2-\sqrt{4s-r^2}C_1\right)\cos\left(\frac{\sqrt{4s-r^2}}{2}\xi\right)}
		{C_1\sin\left(\frac{\sqrt{4s-r^2}}{2}\xi\right)
		+C_2\cos\left(\frac{\sqrt{4s-r^2}}{2}\xi\right)}
	\right\}
	&\qquad (D<0)
	\\
	\quad
	-\frac{rC_1\xi+\left(rC_2-2C_1\right)}{2\left(C_1\xi+C_2\right)}
	&\qquad (D=0)
	\\
	\quad
	-\frac{1}{2}
	\left\{
	\frac{\left(rC_1-\sqrt{r^2-4s}C_2\right)\sinh\left(\frac{\sqrt{r^2-4s}}{2}\xi\right)
		+\left(rC_2-\sqrt{r^2-4s}C_1\right)\cosh\left(\frac{\sqrt{r^2-4s}}{2}\xi\right)}
		{C_1\sinh\left(\frac{\sqrt{r^2-4s}}{2}\xi\right)
		+C_2\cosh\left(\frac{\sqrt{r^2-4s}}{2}\xi\right)}
	\right\}
	&\qquad (D>0)
\end{cases}
\ee
This property enables to translate partial differential equations
into algebraic relations for parameters to be determined by initial conditions.


\end{document}